\documentclass[11pt, showpacs]{article}
\pdfoutput=1
\usepackage{amsmath,amssymb,amsfonts,bm,marginnote,cite,graphicx,slashed}
\usepackage[colorlinks=true, pdfstartview=FitV, linkcolor=black, citecolor=black, urlcolor=black]{hyperref}
\usepackage[usenames,dvipsnames]{xcolor}

\usepackage{epstopdf}
\usepackage[justification=centering]{caption}
\usepackage{subcaption}
\usepackage{dcolumn}
\usepackage{bbm}
\usepackage{amscd}
\usepackage{mathrsfs}
\usepackage{dsfont}
\usepackage{comment}
\usepackage{setspace}
\usepackage[margin=0.75in]{geometry}
\usepackage{tensor}
\usepackage{mathrsfs}
\usepackage{cancel}
\usepackage{hyperref}
\usepackage{empheq}
\usepackage{youngtab}
\usepackage{empheq}
\usepackage{MnSymbol}
\usepackage{float}
\usepackage[all]{xy}

\DeclareMathAlphabet{\mathbbold}{U}{bbold}{m}{n}

\edef\marginnotetextwidth{\the\textwidth}

\parskip=\baselineskip

\newcommand{\thistitle}{
Interface Contributions to Topological Entanglement in Abelian Chern-Simons Theory 

	}

\newcommand{\addressuiuc}{
	Department of Physics, University of Illinois,
 	1110 West Green St., Urbana IL 61801, U.S.A.
	}
	
\newcommand{\addresspenn}{David Rittenhouse Laboratory, University of Pennsylvania,
	209 S. 33rd Street, Philadelphia PA 19104, U.S.A.
	}
	
\newcommand{\addressKITP}{Kavli Institute for Theoretical Physics, University of California, 
	Santa Barbara, CA 93106, U.S.A
	}

\newcommand{\ack}[1]{[{\bf Pfft!: {#1}}]}

\newcommand{\be}{\begin{equation}}
\newcommand{\ee}{\end{equation}}
\newcommand{\beq}{\begin{eqnarray}}
\newcommand{\eeq}{\end{eqnarray}}
\newcommand{\bea}{\begin{eqnarray}}
\newcommand{\eea}{\end{eqnarray}}
\newcommand{\beqn}{\begin{eqnarray}}
\newcommand{\eeqn}{\end{eqnarray}}

\newcommand{\bs}{\boldsymbol}

\newcommand{\cI}{\mathcal I}
\newcommand{\cJ}{\mathcal J}

\def\N{\nabla}
\def\pa{\partial}

\newcommand{\mTr}{\mathrm{Tr}}
\newcommand{\cQ}{\mathcal{Q}}

\newcommand{\myfig}[3]{
	\begin{figure}[ht]
	\centering
	\includegraphics[width=#2cm]{#1}\caption{\small{#3}}\label{fig:#1}
	\end{figure}
	}

\begin{document}

\title{\thistitle}
\author{
	{Jackson R. Fliss$^1$, Xueda Wen$^{1,2}$, Onkar Parrikar$^3$, Chang-Tse Hsieh$^1$, Bo Han$^1$, }\\{Taylor L. Hughes$^1$, Robert G. Leigh$^1$}\\
	\\
	{$^1$\small \emph{\addressuiuc}}
	\\
	{$^2$\small \emph{\addressKITP}}
	\\
	{$^3$\small \emph{\addresspenn}}
\\}
\date{\today}
\maketitle\thispagestyle{empty}

\begin{abstract}
We study the entanglement entropy between (possibly distinct) topological phases across an interface using an Abelian Chern-Simons description with topological boundary conditions (TBCs) at the interface. From a microscopic point of view, these TBCs correspond to turning on particular gapping interactions between the edge modes across the interface. However, in studying entanglement in the continuum Chern-Simons description, we must confront the problem of non-factorization of the Hilbert space, which is a standard property of gauge theories. We carefully define the entanglement entropy by using an extended Hilbert space construction directly in the continuum theory. We show how a given TBC isolates a corresponding gauge invariant state in the extended Hilbert space, and  hence compute the resulting entanglement entropy. We find that the sub-leading correction to the area law remains universal, but depends on the choice of topological boundary conditions. This agrees with the microscopic calculation of \cite{Cano:2014pya}. Additionally, we provide a replica path integral calculation for the entropy. In the case when the topological phases across the interface are taken to be identical, our construction gives a novel explanation of the equivalence between the left-right entanglement of (1+1)d Ishibashi states and the spatial entanglement of (2+1)d topological phases.

\end{abstract}

\newpage

\section{Introduction}

 A chiral topological phase of matter with a boundary is host to gapless boundary modes.  The gapless modes often provide a fingerprint of the bulk topologically ordered phase and can provide universal, measurable phenomena in real material samples.  Heuristically the boundary modes signal a change in topological order from the bulk topological phase to the external vacuum with trivial topological order. An interesting question arises at more complicated heterointerfaces between two topological phases: how do the boundary modes reorganize themselves in the composite system?  In the case where the two phases are topologically identical,  the gapless boundary modes can be ``erased" along the seam via interactions that introduce a gap in the boundary modes, and in this sense make them invisible in the low energy effective theory. Of equal interest is the physics involved in gluing together {\it distinct} topological phases. In some cases one may find that the gapless heterointerface modes \emph{must} persist even when the gapless modes are coupled across the interface with interactions, while in other cases the gapless modes can be unstable to gap formation. It is the latter case in which we are interested.

It is known that the set of gapping interactions that can glue (gap out) the boundary modes for two given topological phases is not unique.  Subject to algebraic constraints, there can exist many choices for gapping interactions.  These classes of gapping interactions were studied in the context of quantum entanglement in Ref. \cite{Cano:2014pya} using `coupled wire' constructions of Abelian topological phases.  There it was shown that the choice of gapping interaction can leave an imprint on the bipartite entanglement spectrum and entropy when the entangling cut is taken along the gapped heterointerface. Interestingly, in this sense the low energy physics should remember the heterointerface, even though it is gapped. 

Explicitly, it was found that the choice of gapping interactions can modify the low-energy entanglement spectrum and the sub-leading correction to the area law in the entanglement entropy. The latter effect is perhaps most surprising because the constant, sub-leading correction is known to be a universal, topological quantity \cite{Levin:2006zz, Kitaev:2005dm}. The calculations that predict these effects all rely on coupled-wire constructions, and while such constructions are theoretically convenient, there are some limitations in their description. For example, they are discretized in at least one spatial direction, i.e., the system is made from discretized wires/strips, and translation symmetry is implicitly assumed parallel to the wires. These two issues limit the types of entanglement cuts and spatial geometries that can be simply handled. In principle, lacking any corroborating analytic or numeric calculations, it is not clear if all of the conclusions of \cite{Cano:2014pya} are independent of the coupled wire model description. Here we seek to support and extend these results of \cite{Cano:2014pya} using a more generic field theoretical approach.

The goal of this paper is to revisit the question of entanglement along heterointerfaces from the point of view of the bulk topological theory in the continuum.  Indeed, the effective low energy physics at the boundary of a topological phase is mirrored by a bulk topological field theory, through anomaly inflow.  From the point of view of entanglement in the bulk theory there are two questions that become immediately relevant.
First, when we have heterointerfaces of topological phases, it is natural to ask how to address the gapping physics in the continuum.
  For the Chern-Simons theory describing Fractional Quantum Hall (FQH) states, the natural answer to this question involves a set of prescribed conditions for matching the gauge fields living on either side of the interface. Such conditions might be termed {\it interface conditions}. However, in the case where the 
  spaces on either side of the interface are homeomorphic,  by regarding the space hosting the two phases as the Schottky double of a single topological phase, the matching conditions can be thought of as boundary conditions, and this is the language we will use in this paper.  Generic boundary conditions available to Chern-Simons theory will typically ``break" the topological nature, i.e., they will introduce a complex structure on the boundary in question.  This, for instance, is necessary for defining the chirality of the induced gapless boundary modes for a single topological phase.  
  However, as an alternative, it is possible to choose {\it topological boundary conditions}, which obviate the need for a complex structure.  We will argue that these boundary conditions are appropriate for the context of gapped interfaces, at least as far as the low energy physics is concerned. 
More precisely, at least perturbatively, and as long as the gap does not close, the addition of non-topological boundary terms are expected to modify only the quantitative details of the low-energy theory.

Topological boundary conditions (TBCs) have been discussed and classified in previous literature.  Generically, the choice of TBCs is itself not unique and depends on certain algebraic properties of the $K$-matrix.  Additionally, since TBCs are playing the same role as gapping interactions for the topological field theory (that is, they glue two theories together), it is perhaps unsurprising then that these algebraic criteria are equivalent to those classifying gapping interactions.  These criteria were discussed at length in Ref. \cite{Kapustin:2010hk} where it was emphasized that these boundary conditions isolate a Lagrangian subspace of the $K$-matrix and so pick a polarization for states on the interface.  It was additionally pointed out in Ref. \cite{Wang:2013yta} that TBCs are equivalent to anomaly matching: the unbroken gauge group on the boundary remains anomaly-free.

The second question to address is how to define entanglement in the continuum gauge theory.  As has been recently understood, the Hilbert spaces of theories with gauge invariance generically do not admit tensor-factorization of spatial regions due to the enforcement of non-local constraints \cite{Buividovich:2008gq, Donnelly:2011hn}.  This is mirrored by the fact that the set of gauge invariant operators are generated by Wilson loops which are inherently non-local.  Several proposed attempts to define bipartite entanglement in gauge theories have been studied in the recent literature \cite{Soni:2015yga, Donnelly:2011hn, Donnelly:2014gva, Donnelly:2016auv, Ghosh:2015iwa, Casini:2013rba, Casini:2014aia, Donnelly:2014fua, Donnelly:2015hxa} and can be classified into two approaches: the \emph{algebraic} approach and the \emph{extended Hilbert space} approach.  The former focuses on the definition of the reduced density matrix as living in an operator algebra.  There the inability of the Hilbert space to factorize is mirrored by the existence of a non-trivial center of the algebra associated to a subspace; the reduced density matrix is block-diagonalized with respect to this center and entanglement is computed in each block.  The latter approach embeds the physical Hilbert space in a larger, factorizable Hilbert space.  Generically this space contains states that do not obey gauge invariance and the gauge invariant state must be identified by the application of constraints.  Once identified, the state can be reduced and the entanglement computed.  It is worth mentioning that while both approaches have well-defined and controlled procedures on the lattice, the extension to continuum gauge theories is subtle.

In this paper, we will confine our discussion to Abelian Chern-Simons theories with gauge group $U(1)^N$. Our main interest will be in studying the effect of topological boundary conditions on the entanglement entropy across an interface.  We will discuss how to unambiguously embed the physical, gauge-invariant Hilbert space of Chern-Simons theory in the presence of such an interface into a tensor-factorizable \emph{extended Hilbert space}.  Remarkably, the physical states (thought of as living inside the extended Hilbert space) satisfy a generalized Ishibashi condition specific to the choice of topological boundary conditions, which allows a straightforward computation of the entanglement entropy. This offers a novel explanation of the known equivalence of spatial entanglement in Chern-Simons theory and the left-right entanglement of Ishibashi states\cite{PhysRevLett.108.196402, Das:2015oha, Wen:2016snr}, and offers a connection of our problem to the recent papers addressing the entanglement of bosonic CFTs across topological interfaces \cite{Fuchs:2007tx, Sakai:2008tt, Gutperle:2015kmw, Brehm:2015plf}.  Additionally, we also give a replica path integral calculation of the entanglement entropy.  We pursue this in two ways -- the first is via a direct path integral calculation involving the replica trick within Chern-Simons theory.  Passing to field variables obeying the TBCs within the path integral, we find that the replica computation reduces to a familiar Chern-Simons path integral but with an \emph{effective $K$-matrix.}  The second involves introducing a regulator surface enveloping the entanglement cut and reducing the path integral to a transition amplitude between CFT boundary states living at the intersection of the regulator surface with the heterointerface.  The results obtained using all the above techniques of course agree with each other, and also with the microscopic calculations of \cite{Cano:2014pya}. This shows that the entanglement entropy depends explicitly upon the choice of TBCs, or equivalently from a microscopic point of view on the choice of gapping interactions between the edge-modes across the interface,  but that this is nevertheless a universal feature which can be reproduced from the effective topological field theory description.  

The rest of this paper is organized as follows:  in Section \ref{ClassSympBCs} we begin with a brief classical discussion of interfaces in Abelian Chern-Simons theory with topological boundary conditions, and their relation with the gapping interactions studied in \cite{Cano:2014pya}.  We then consider the quantum version of these interfaces in Section \ref{QuantInt&TEE} and present the extended Hilbert space calculation of the entanglement entropy across such interfaces. In Section \ref{sec4}, we reproduce the same result using the replica path integral in the two  ways discussed above.  Finally, we have a short conclusion disussing future directions and two appendices.

\section{Classical Interfaces and Topological Boundary Conditions}\label{ClassSympBCs}
In this section, we discuss classical interfaces in Abelian Chern-Simons theory with the gauge group $U(1)^N$. Recall that the action of $U(1)^N$ Chern-Simons theory on a 3-manifold $M$ is given by
\beq
S_{CS} = \frac{1}{4\pi} \int_M K^{IJ} A_I \wedge dA_{J}\label{eq:CSeq1}
\eeq
where $I=1, \cdots, N$ and $K^{IJ}$ is a symmetric integral matrix of rank $N$ called the level matrix. By an \emph{interface}, we mean a codimension-one surface $\Sigma$ in $M$, with different $K$-matrices on either side of $\Sigma$ (see Figure \ref{fig:fig1.pdf}). In particular the Chern-Simons theories on either side of $\Sigma$ differ in their actions, 
namely in their $K$-matrices. We denote the gauge fields on the left and right of the interface by $A^{(L)}_{I}$ and $A^{(R)}_I$, and the respective $K$-matrices by $K^{(L)}$ and $K^{(R)}$. 
\myfig{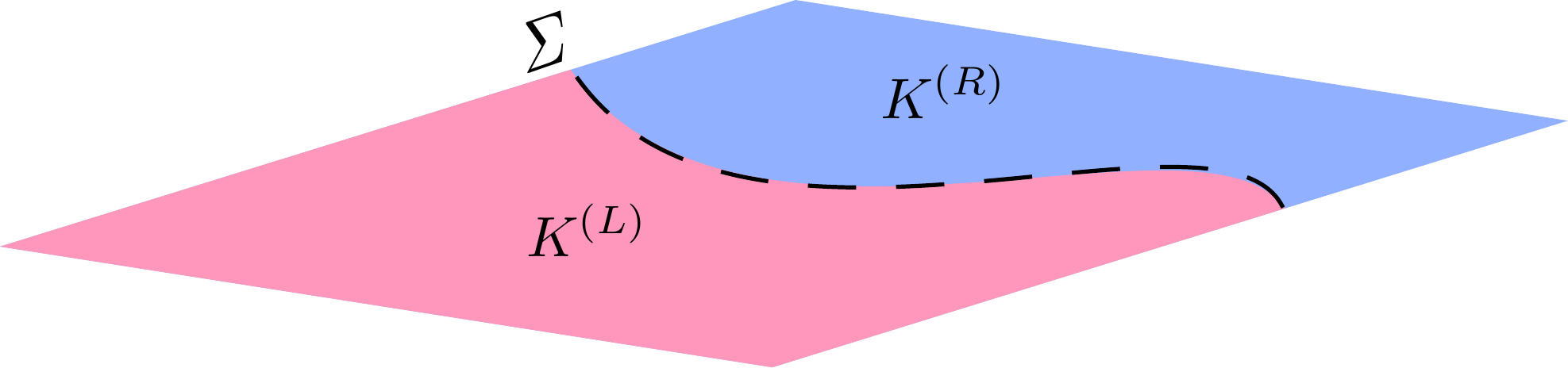}{10}{\textsf{Two topological phases separated a codimension one defect $\Sigma$. The time dimension has been suppressed here.}}
 We denote the space-time as $M=\mathcal M_L\cup_\Sigma \mathcal M_R$ and write the action as 
\begin{equation}
S_{CS}=\frac{{K^{(L)}}^{IJ}}{4\pi}\int_{\mathcal M_L}A^{(L)}_I\wedge dA^{(L)}_J+\frac{{K^{(R)}}^{IJ}}{4\pi}\int_{\mathcal M_R}A^{(R)}_I\wedge dA^{(R)}_J+S_{\Sigma}(A_L,A_R).
\end{equation}
Here $S_{\Sigma}$ denotes additional boundary/interface terms with support on $\Sigma$ which one might possibly add; 
we will make some comments on the role of these boundary terms shortly.  In the present section, we will consider the above theory from a \emph{classical} point of view, focusing on interface boundary conditions; we will then revisit interfaces from a quantum point of view in the next section. There are a number of generalizations that we might make, including the discussion of non-Abelian Chern-Simons theories, but we will leave these to future work.

Consistent boundary conditions on $\Sigma$ are determined by ensuring that the \emph{symplectic structure} 
 is continuous across $\Sigma$.  We take a brief detour to explain what this means. The variation of the Chern-Simons action \eqref{eq:CSeq1} on a general 3-manifold $M$ with boundary $\pa M$ is given by
\beq
\delta S_{CS}=\frac{1}{2\pi} \int_{M}K^{IJ} \delta A_I\wedge dA_J -\frac{1}{4\pi}\int_{\pa M} K^{IJ}a_I\wedge \delta a_J+\delta S_{\pa M}
\eeq
where $a$ is the connection on $\pa M$ induced from $M$, and the last term above comes from the variation of any potential boundary terms. The classical equations of motion are given by $K^{IJ}dA_J=0$. We regard the variation of the action taken on-shell (o.s.) as a 1-form on the field space of classical solutions\footnote{That is we regard $\delta$ as a field space differential and these should be regarded formally as anticommuting. In the text, we explicitly denote the wedge products of forms on $M$, while leaving the antisymmetrization of forms on field space implicit. Thus, for example, eq. (\ref{CSSymp2Form}) is non-zero for a {\it symmetric} $K$-matrix.};  we will denote this as the \emph{canonical symplectic 1-form}, $\bs \Theta$.  For the present action it is defined as
\beq
\bs\Theta  = \delta S_{CS}\Big|_{o.s.} = -\frac{1}{4\pi}\int_{\pa M} K^{IJ}a_I\wedge \delta a_J+\delta S_{\pa M}.
\eeq
The \emph{symplectic 2-form} is the differential of the canonical 1-form:
\begin{equation}\label{CSSymp2Form}
\bs \Omega= \delta \bs \Theta=-\frac{1}{4\pi}\int_{\pa M}\left(K^{IJ}\delta a_I\wedge \delta a_J\right).
\end{equation}
Because $S_{\pa M}$ adds an exact form to the canonical 1-form, the symplectic 2-form is unaffected by its presence.  Because $K$ is non-degenerate, $\Omega$ promotes the classical phase space to a symplectic vector space. Although generically this vector space will be infinite dimensional, Chern-Simons theory provides us with many cases in which it is finite (e.g., when the gauge group is compact and $\pa M$ is closed and compact \cite{Elitzur:1989nr}); the process of choosing boundary conditions amounts to finding a half-dimensional subspace upon which the symplectic form vanishes.  For example, standard boundary conditions in Chern-Simons amount to fixing some component of $a$ on the boundary.  The role of $S_{\pa M}$ then is to ensure that $\bs \Theta$ vanishes when restricted to fields obeying this boundary condition\footnote{This is equivalent to the continuity across the cut of the symplectic one-form on a family of hypersurfaces $\Sigma_t$ parallel to $\Sigma$.}; alternatively this can be thought of 
ensuring a well-defined variational principle.  We will refer to this as putting $\bs \Theta$ in \emph{canonical form.}  Fixing a component of $a$ generically involves the introduction of a boundary term that either breaks diffeomorphism invariance on $\pa M$, or introduces a metric structure on $\pa M$.  The standard boundary term for this is
\begin{equation}\label{nontopBC}
S_{\pa M}=\frac{1}{4\pi}\int_{\pa M} V^{IJ}a_I\wedge \ast a_J
\end{equation}
where $\ast$ is the Hodge star for a Riemannian metric on $\pa M,$ and $V^{IJ}$ is taken to be a symmetric, positive-definite matrix.  Suitable choices for $V$ enforce the fixing of either the holomorphic or antiholomorphic (with respect to the orthonormal coordinates of the metric) component of $a$.  When reducing a Chern-Simons path integral to that of a chiral Wess-Zumino-Witten (WZW) theory on $\pa M$, such a boundary term introduces dynamics, i.e., a non-vanishing Hamiltonian, as we discuss further in Section \ref{sectinhomoishi}.

Alternatively, returning to \eqref{CSSymp2Form}, for suitable even-dimensional $K$-matrices, we can look for vanishing subspaces of $K$ itself. As we will see, these boundary conditions do not require the addition of an extra boundary action and in particular do not require a choice of metric on $\pa M$.  As such they are called \emph{topological boundary conditions} \cite{Kapustin:2010hk}. In a completely generic physical context, one would expect both metric-dependent bulk terms in the action (e.g., a Maxwell term), and additional metric-dependent boundary terms as well. However, we expect that the effect of such terms is to modify inessential details of the gapped boundary/interface theory\footnote{To clarify, it is well known that the degrees of freedom in Maxwell-Chern-Simons theory decouple into a flat connection and a topologically massive gauge field \cite{Andrade:2011sx}.  The former contributes to the topological entanglement entropy,  while the latter adds a massive contribution to the entanglement entropy \cite{Agarwal:2016cir}.}, and that moreover, the study of topological boundary conditions is sufficient for the study of many properties (such as entanglement) of the gapped interfaces in which we are interested.  As such, we will ignore these extra possible terms.

To elaborate on this, let us return to the case of the interface theory which is of interest in the present paper. An equivalent way to think about this theory on $\mathcal M_L\cup_\Sigma \mathcal M_R$ (in the case where $\mathcal M_L$ and $\mathcal M_R$ are topologically equivalent) is to ``fold" the theory along the common boundary, $\Sigma$.\footnote{Of course, in a generic situation we do not mean to require that $\mathcal M_{L,R}$ are homeomorphic. Indeed, they could very well have different topology. Consequently, this discussion can be thought of as applying to a tubular neighbourhood of $\Sigma$, but we will present the material from a simplified point of view.} 
Doing so, we obtain a Chern-Simons theory with gauge group $U(1)^{2N}$ on the space $\mathcal N$ (which is topologically equivalent to $\mathcal M_{L,R}$) with boundary $\pa \mathcal N=\Sigma$. The $K$-matrix of this theory is given by
\beq
\mathbb{K} = K_L\oplus (-K_R).
\eeq 
The signature (number of positive eigenvalues and number of negative eigenvalues) of $\mathbb K$ is $(N,N)$ if, for example, $K_{L,R}$ are each positive definite.  For the rest of this paper we will assume this is the case, although this is not a necessary supposition (removing the assumption only modifies some details of the calculations we present); what is necessary is that the \emph{total signature} (the number of positive eigenvalues minus the number of negative eigenvalues) of $\mathbb K$ is zero.  
Under these conditions, we re-write the action as 
\begin{equation}
S_{CS}=\frac{\mathbb{K}^{\cI \cJ}}{4\pi}\int_{{\mathcal N}}\mathcal A_\cI\wedge d \mathcal A_\cJ
\end{equation}
where
\begin{equation}\label{kmat}
\mathbb{K}=\left(\begin{array}{cc}K^{(L)}&0\\0&-K^{(R)}\end{array}\right),\qquad\qquad\qquad\qquad 
\mathcal A=\left(\begin{array}{c}A^{(L)}\\A^{(R)}\end{array}\right).
\end{equation}
Below we will denote the induced $U(1)^{2N}$ connection on $\Sigma$ as $\mathfrak{a}$. Let us then consider what kind of boundary conditions can be imposed on the field $\mathcal A$.

Reviewing \cite{Kapustin:2010hk}, we will regard $U(1)^{2N}$ as a torus, ${\mathbb T}_\Lambda={\mathbb R}^{2N}/\Lambda$, for $\Lambda\simeq\mathbb Z^{2N}$.  The corresponding Lie algebra will be denoted $\mathfrak t_\Lambda\simeq\Lambda\otimes \mathbb R$. In this language, $\mathbb K$ is an integral symmetric bilinear form on $\Lambda$.  
  The dual lattice $\Lambda^*$ is the set of homomorphisms from $\Lambda$ to the integers that we will denote as the lattice of charges.  It is clear that the image of $\mathbb K$, $\text{Im}\left(\mathbb K\right)$, is contained in $\Lambda^*$.  Basic gauge invariant operators of the theory are constructed from $q\in\Lambda^*$ by Wilson loop operators
  \begin{equation}
  \langle W_q\rangle=\langle\exp\left(i\oint_{C}q(\mathcal A)\right)\rangle=\langle\exp\left(i\oint_Cq^{\cI}\mathcal A_\cI\right)\rangle=\exp\left(-2\pi iq^{\cI}\,\mathbb K^{-1}_{\cI\cJ}\,q^{\cJ}_{total}\right)
  \end{equation}
  where $q_{total}$ is the sum total of additional Wilson loop operators threading $W_q$.\footnote{The expectation value results from the following: the presence of additional Wilson loops operators augment the background equations of motion to $d\mathcal A=-2\pi \sum_m\mathbb K^{-1}\cdot q_m\delta(C_m)$, where $\sum_mq^{\cI}_m=q^{\cI}_{total}$ and $\delta(C_m)$ indicate the flux only has support along the contours threading the original Wilson loop.  Evaluating the holonomy of such a connection produces the above result.}  Thus, 
Wilson loop operators with charges differing by an element of $\text{Im}\left(\mathbb K\right)$ will have identical expectation values, and so it is natural  to work with operators in the quotient $D\equiv\Lambda^*/\text{Im}\left(\mathbb K\right)$\cite{Kapustin:2010hk}.  As discussed previously, the continuity of the symplectic structure across the interface (consistent boundary conditions) now requires 
\beq
\bs\Omega  = -\frac{1}{4\pi}\int_{\Sigma} \mathbb{K}^{\cI \cJ}\delta\mathfrak{a}_\cI\wedge \delta \mathfrak{a}_\cJ = 0
\eeq where $\cI, \cJ= 1,2, \ldots 2N.$
In this paper, we will focus on the class of topological boundary conditions in which  $\mathfrak a$ lies in a \emph{Lagrangian subspace} of $\mathbb K$ \cite{Kapustin:2010hk, Wang:2013yta, Levin:2013gaa} . A subspace $\mathfrak{t}_0 \subset \mathfrak{t}_{\Lambda}$ is called Lagrangian with respect to $\mathbb{K}$ if
\beq
v^\cI \mathbb{K}_{\cI \cJ} w^\cJ = 0,\;\;\;\; \forall v,w \in \mathfrak{t}_0,
\eeq
 and has a dimension that is half the rank of $\mathbb K.$ Such a subspace exists only if the total signature of $\mathbb K$ is zero \cite{Kapustin:2010hk}.  From equation \eqref{kmat}, this means that the signature of $K^{(L)}$ must equal the signature of $K^{(R)}$.  As mentioned above, we will implicitly assume that both $K^{(L)}$ and $K^{(R)}$ are positive definite and so must have the same rank if the total signature is to vanish.  At the level of the canonical 1-form, we see that $\bs\Theta$ is canonical without the addition of a boundary action:
\beq
\bs\Theta  = -\frac{1}{4\pi}\int_{\Sigma} \mathbb{K}^{\cI \cJ}\mathfrak{a}_\cI\wedge \delta \mathfrak{a}_\cJ = 0.
\eeq

The restriction of $\mathfrak a$ to a Lagrangian subspace of $\mathbb K$ means that it takes values in a subalgebra $\mathfrak t_0\subset\mathfrak t_\Lambda$ whose dimension as a Lie algebra is half that of $\mathfrak t_\Lambda$.   The restriction to the subalgebra $\mathfrak t_0$ has another important interpretation: infinitesimal $U(1)^N$ transformations lying in $\mathfrak t_0$ have a vanishing inner product with the canonical 1-form and so there are no dynamical degrees of freedom carrying charge with respect to this group. That is to say $\mathfrak t_0$ generates an \emph{unbroken} $U(1)^N$ of true gauge transformations.  In the context of the unfolded theory, we see that TBCs then ensure that a particular linear combination of fields remain gauge invariant across $\Sigma$.
\begin{figure}[h!]
  \centering
  \begin{tabular}{ c c c }
    \includegraphics[width=.25\textwidth]{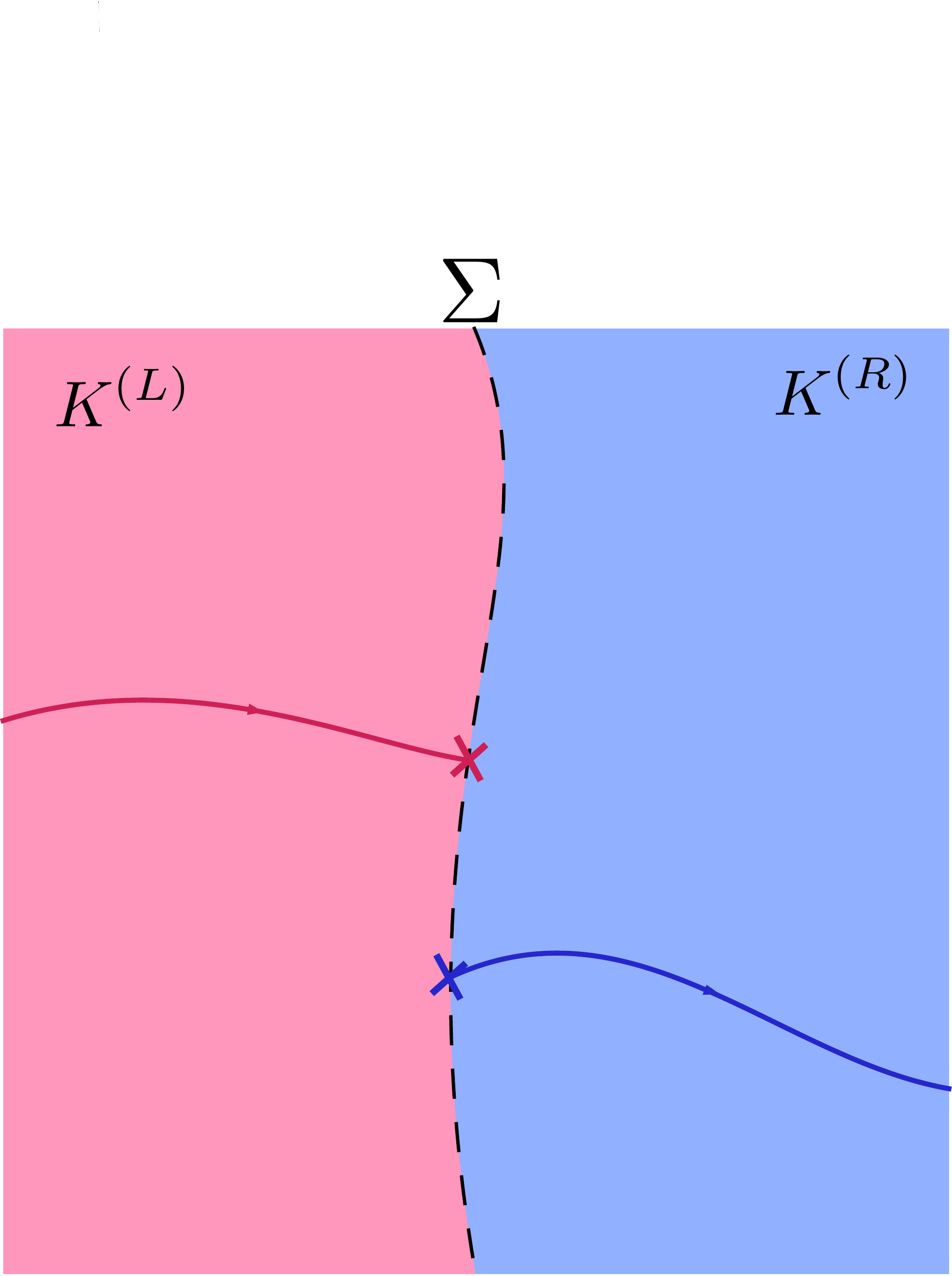} & \hspace{3 cm} & \includegraphics[width=.25\textwidth]{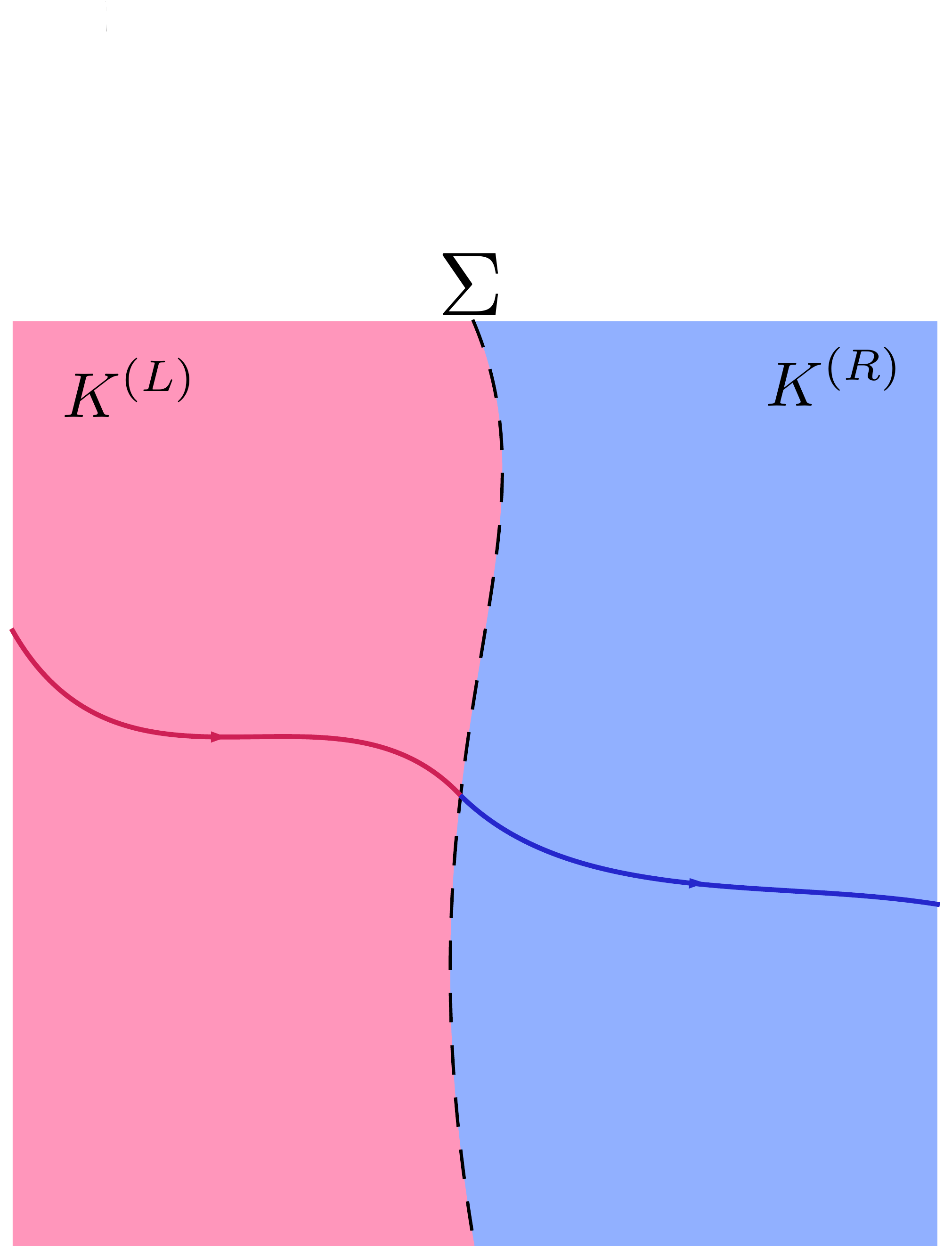}
    \end{tabular}
      \caption{\small{{(Left) Generically an interface will support global $U(1)^N$ charges (depicted here as Wilson lines ending on $\Sigma$.  (Right) The TBCs provide an identification of the gauge group across $\Sigma$ and therefore describe Wilson lines that can permeate the interface. }}}
      \label{WLgluefig}
\end{figure}\\\\
We denote the injection of $\mathfrak t_0$ into $\mathfrak t_\Lambda$ as $\mathbb P$.  We then have
\begin{equation}\label{injPK}
\mathbb P^T\cdot\mathbb K\cdot \mathbb P=0.
\end{equation}
The unbroken Lie algebra can be described as $\mathfrak t_0\simeq\Lambda_0\otimes\mathbb R$ for a lattice $\Lambda_0\subset\Lambda$ such that the unbroken group is a torus $\mathbb T_0=\mathbb R^N/\Lambda_0$. We will sometimes call this the \emph{restricted lattice}.  In this context, $\mathbb P$ is an injection of $\Lambda_0$ into $\Lambda$ and $\mathbb P^T$ is a surjection from the lattice of charges $\Lambda^*$ to a sublattice of boundary charges $\Lambda^*_0$.  We will refer to this as the \emph{restricted dual lattice.}  Given a basis for $\Lambda$, $\mathbb P$ is an integral matrix with $2N$ rows and $N$ columns.  In the current basis let us 
choose 
 \[\mathbb P=\left(\begin{array}{c}v^{(L)}\\-v^{(R)}\end{array}\right)\] for $N\times N$ integer matrices $v^{(L,R)}$, in terms of which (\ref{injPK}) becomes
\begin{equation}\label{injcond}
{v^{(L)}}^T\cdot K^{(L)}\cdot v^{(L)}-{v^{(R)}}^T\cdot K^{(R)}\cdot v^{(R)}=0.
\end{equation}
We will refer to \eqref{injcond} as the \emph{classical gluing condition.}  Then in order for topological boundary conditions to exist between the two theories, $K^{(L)}$ and $K^{(R)}$ must allow for integral solutions to \eqref{injcond}, a significantly non-trivial condition.  However, if such a solution exists, then an infinite number of solutions exist: for example, multiplying $v^{(L)}$ and $v^{(R)}$ by the same integer will also solve \eqref{injcond}.  In this paper, we project to a minimal set of solutions by requiring $\mathbb P$ to be \emph{primitive} \cite{Levin:2009orm}.  That is, expressing $\mathbb P$ as a $2N\times N$ integral matrix, we require that the ${\tiny \left(\begin{array}{c}2N\\N\end{array}\right)}$ possible $N\times N$ minors have a gcd of 1.  We give a geometric interpretation of this condition in Appendix \ref{sectGeomPrim}, but for a discussion of primitivity in the condensed matter context, see \cite{Cano:2014pya, Wang:2013vna, Levin:2011hq, Levin:2012ta}.

Note that in the unfolded theory, \eqref{injcond} tells us that particular linear combinations of the connections can permeate the interface.  These connections then see an effective $K$-matrix which is continuous across the interface:
\begin{equation}
K_{eff}\equiv {v^{(L)}}^T\cdot K^{(L)}\cdot v^{(L)}={v^{(R)}}^T\cdot K^{(R)}\cdot v^{(R)}.
\end{equation}
In the present context, $K^{(L/R)}$ have the same rank, and so $v^{(L/R)}$ are square matrices.  The above equation then details what linear combinations of the gauge field permeates $\Sigma$ so that it remains anomaly free under gauge transformations in $\mathbb T_0$.  However, as we have previously emphasized, it is not actually necessary to take $K^{(L,R)}$ to have the same rank, but only that $\mathbb{K}$ have zero total signature.  When the ranks of the $K$-matrices differ then $v^{(L/R)}$ no longer have to be square and so only a subspace of charges can permeate $\Sigma$.  

\subsection{Examples}

For the sake of pedagogy, let us examine some examples of gappable interfaces.

\subsubsection{$K^{(L)}=(k_L)$, $K^{(R)}=(k_R)$}
For a first example $K^{(L,R)}$ are positive integer $1\times1$ ``matrices", $(k_{L,R})$.  In this case, the gluing condition has us looking for integer solutions $\{v^{(L)}, v^{(R)}\}$ to
\begin{equation}\label{ex1eq}
{v^{(L)}}^2k_L={v^{(R)}}^2k_R.
\end{equation}
Primitivity requires that $v^{(L)}$ and $v^{(R)}$ be relatively prime.  Let $k=gcd[k_L,k_R]$ and write $k_{L,R}=\kappa_{L,R}\,k$ with $\kappa_L$ and $\kappa_R$ relatively prime integers.  We find that integer solutions to \eqref{ex1eq} only exist if $\kappa_{L,R}$ are perfect squares ($\kappa_{L,R}=n_{L,R}^2$ for some integers $n_{L,R}$).  Then there are exactly four solutions:
\begin{equation}
v^{(L)}=\pm n_R\qquad\qquad\qquad v^{(R)}=\pm n_L.
\end{equation}
One can readily check for $\mathbb P^T=\left(\pm n_R, \pm n_L\right)$ and $\mathbb K=(n_L^2\,k)\oplus (-n_R^2\,k)$ that $\mathbb P^T\cdot \mathbb K\cdot \mathbb P=0,$ and so this also defines a Lagrangian subspace for $K^{(L)}\oplus(-K^{(R)})$.  The effective $K$-matrix for this interface is
\begin{equation}
K_{eff}={v^{(L)}}^2K^{(L)}={v^{(R)}}^2K^{(R)}=k\,n_L^2\,n_R^2.
\end{equation}
This example is particularly instructive because we see that the gluing condition not only determines the matrices $v^{(L)}$ and $v^{(R)},$ but also restricts the set of $K$-matrices.

\subsubsection{$K^{(L)}=K^{(R)}=K$}

As a second example, we illustrate the fact that $K$ can be glued to itself in not only the trivial manner (e.g., the gauge field being continuous across the interface, $a_L=a_R$, corresponds to the solution $v^{(L)}=-v^{(R)}=1_{N\times N}$), but in fact, in multiple ways.  Thus, even in this homogeneous case, the choice of boundary conditions is far from unique.  Taking the determinant of the gluing condition, we have
\begin{equation}
\text{det}(v^{(L)})^2=\text{det}(v^{(R)})^2.
\end{equation}
Depending on the details of $K$, 
we may have several solutions beyond the identity matrices that solve the gluing condition.  As an explicit example,
\begin{equation}
K=\left(\begin{array}{cc}k&0\\0&k(m^2-n^2)\end{array}\right)\qquad\qquad v^{(L)}=\left(\begin{array}{cc}m&n\\0&1\end{array}\right)\qquad\qquad  v^{(R)}=\left(\begin{array}{cc}-n&-m\\1&0\end{array}\right)
\end{equation}
for integers $k, m, n$ and $m^2\neq n^2$, solves the gluing condition.  One can easily check that the minors of $\mathbb P$ are $\{m,-m^2+n^2, -n, n, -1, m\}$ and so this solution is also primitive.  The effective $K$-matrix is 
\begin{equation}
K_{eff}=\left(\begin{array}{cc}k\,m^2&k\,mn\\k\,mn&k\,m^2\end{array}\right).
\end{equation} Interestingly the subleading correction to the area law in the entanglement entropy depends on the topological boundary conditions.  The initial $K$-matrix has a determinant $\gamma=k^2(m^2-n^2)$ whereas the effective matrix has $\gamma_{eff}=k^2 m^2(m^2-n^2).$ As we will see, the subleading correction to the area law is modified from $-\frac{1}{2}\log[\gamma]$ when trivial boundary conditions are chosen to $-\frac{1}{2}\log[\gamma_{eff}]$ when the more complicated topological boundary conditions are chosen.
For more examples of $K^{(L)}=K^{(R)}$ gapped interfaces, see \cite{Cano:2014pya}.

\bigskip
\subsection{Comments on Topological Boundary Conditions and the Connection to the Coupled Wire Construction}
Before moving on to the discussion of entanglement, we remark that in some instances it will be convenient to describe topological boundary conditions by the injection of the complementary space of $\mathfrak t_{0}$.  We will call this injection $\mathbb M:\mathfrak t_0^c\hookrightarrow \mathfrak t_\Lambda$ and so the fields on $\Sigma$ can be equivalently characterized by the kernel of ${\mathbb M}^T$:
\begin{equation}\label{kerBC}
\mathfrak a\in\mathfrak t_0\qquad\Rightarrow\qquad{\mathbb M}^T\cdot \mathfrak a=0.
\end{equation}
Given the block basis for $\mathbb P$, we can write ${\mathbb M}$ out explicitly as
\begin{equation}
{\mathbb M}=\left(\begin{array}{c}M_L\\ M_R\end{array}\right)=\left(\begin{array}{c}K^{(L)}\cdot v^{(L)}\\K^{(R)}\cdot v^{(R)}\end{array}\right).
\end{equation} 
It is easy to verify that vectors in the pre-image of $\mathbb P$ lie in the kernel of ${\mathbb M}^T$.  Suppose we have a primitive solution to \eqref{injcond}, and let $a$ be a $\mathfrak t_0$-valued connection on $\Sigma$ in the pre-image of the injection $\mathbb P$:
\begin{equation}
\mathfrak a=\left(\begin{array}{c}a^{(L)}\\a^{(R)}\end{array}\right)=\mathbb P\cdot a=\left(\begin{array}{c}v^{(L)}\cdot a\\-v^{(R)}\cdot a\end{array}\right).
\end{equation}
Then
\begin{equation}
{\mathbb M}^T\cdot \mathfrak a=\left(\begin{array}{cc}{v^{(L)}}^T\cdot K^{(L)},&{v^{(R)}}^T\cdot K^{(R)}\end{array}\right)\left(\begin{array}{c}v^{(L)}\cdot a\\-v^{(R)}\cdot a\end{array}\right)=\left({v^{(L)}}^T\cdot K^{(L)}\cdot v^{(L)}-{v^{(R)}}^T\cdot K^{(R)}\cdot v^{(R)}\right)\cdot a=0.
\end{equation}  

Although it is known in the condensed matter literature that TBCs are equivalent to primitive gapping vectors \cite{Wang:2013yta}, let us offer an intuitive picture of this relation in terms of the gapping interactions studied in \cite{Cano:2014pya}.  We first notice that in terms of the matrix ${\mathbb M}$ in the block basis, the equation to be solved for the boundary to support topological boundary conditions is
\begin{equation}\label{Mbcs}
{M_L}^T\cdot \left(K^{(L)}\right)^{-1}\cdot M_L-{M_R}^T\cdot \left(K^{(R)}\right)^{-1}\cdot M_R=0.
\end{equation}
which is precisely the commensurability condition 
of gapping vectors as usually presented.\cite{Cano:2014pya, Haldane:1995xgi}

\myfig{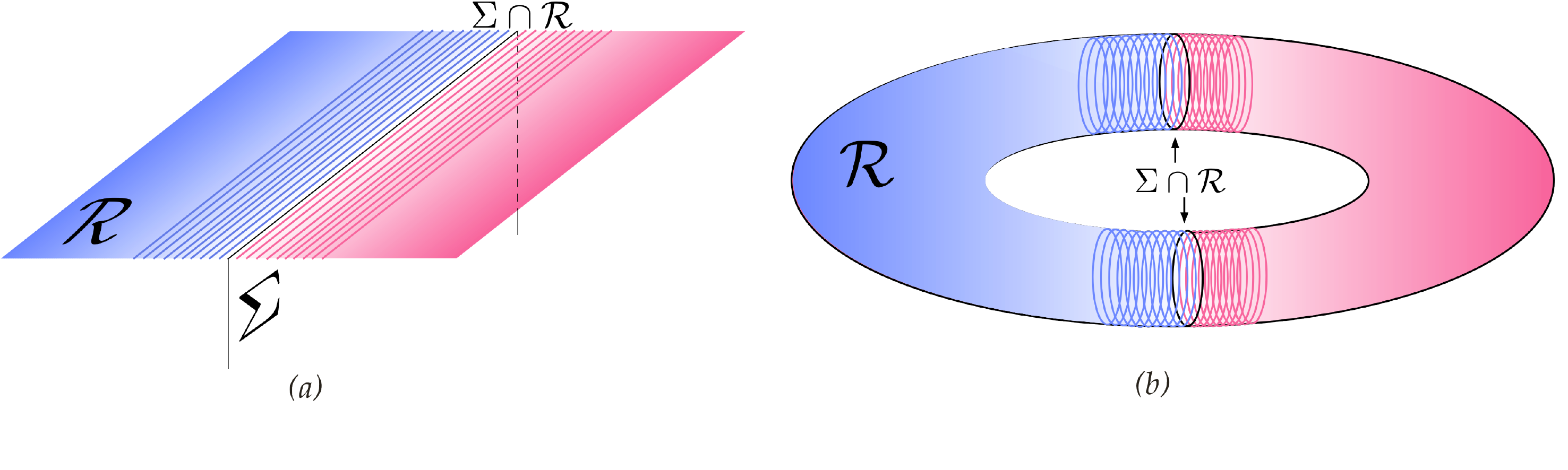}{15}{\textsf{(a) The hypersurface $\mathcal R$ intersects the interface $\Sigma$ transversely.  In the ``coupled wire construction," $\mathcal R$ is foliated by one dimensional wires each hosting a bosonic theory.  In the continuum, $\mathcal R$ supports a connection that breaks up into components normal and tangent to $\Sigma$. 
(b) $\mathcal R$ can possess noncontractible cycles and correspondingly the bosonic theory will contain winding modes.}}

To illustrate the significance of $M_{L,R}$, let us define a hypersurface $\mathcal R$ that intersects the surface $\Sigma$ transversely. For example, we can regard $\mathcal R$ as a  constant-time hypersurface, as shown in Fig. \ref{fig:wiresboth.pdf}(a). 
Relating this to the discussion in \cite{Cano:2014pya}, one can think of $\mathcal R$ as discretized into a family of wires, or strips, running parallel to $\Sigma$ as a model for the gapped topological phase \cite{mukhopadhyay2001,kane2002,teo2014,Cano:2014pya}. 
 In the continuum, we proceed by breaking the connection into components normal to (i.e., those that pull back to zero) and tangent to $\Sigma$. The normal component of $A$ acts as a Lagrange multiplier, constraining the components of the connection in the directions parallel to $\Sigma$ to be flat.  We then write\footnote{By the map $i$, we denote the inclusion map of a wire into $M$, taken on either side of $\Sigma$. Since $U(1)^N$ is not simply connected, the definition of $\phi^{(L,R)}$ should be taken to apply locally, or equivalently, that $\phi^{(L,R)}$ is not required to be single-valued.} $a^{(L,R)}=i_*A^{(L,R)}=d\phi^{(L,R)}$.  The action in terms of these pure gauge modes is a total derivative and so the path integral is then over the $U(1)^{2N}$-valued fields living on $\mathcal R$ (see, for instance, \cite{Elitzur:1989nr}).  This is the standard reduction of Abelian Chern-Simons theory to $U(1)$ Wess-Zumino-Witten (WZW) on $\mathcal R$.

The TBCs on $a^{(L,R)}$ will impose boundary conditions on the field $\phi^{(L,R)}$ on the interface $\mathcal R\cap\Sigma$ (or, from the coupled wire point of view, impose conditions on the wires on either side of $\Sigma$).  Above we saw that $\mathbb P$ embeds the unbroken boundary Lie algebra, $\mathfrak t_0$, into the original Lie algebra, $\mathfrak t_\Lambda$.  This embedding is in the kernel of ${\mathbb M^T}$ (c.f. eq. (\ref{kerBC})).  Then, up to a constant shift, 
 the pure gauge modes of $A^{(L,R)}$
 are related via
\begin{equation}\label{WZWbcs}
M_L^T\cdot\phi^{(L)}+M_R^T\cdot\phi^{(R)}=0.
\end{equation}
If $\mathcal R$ contains non-contractible cycles (as for example in the situation sketched in Fig. \ref{fig:wiresboth.pdf}(b)), then $\phi^{(L,R)}$ could have non-trivial integer windings: $\phi_I^{(L,R)}(x+2\pi)=\phi_I^{(L,R)}(x)+2\pi \mathcal P_I^{(L,R)}$.  However the boundary conditions \eqref{WZWbcs} tell us that these windings must lie in the \emph{restricted lattice}, $\Lambda_0$: 
\begin{equation}\label{restrictedlattice}
\left(\begin{array}{cc}\mathcal P^{(L)}\\\mathcal P^{(R)}\end{array}\right)\in\Lambda_0\qquad\Rightarrow\qquad M_L^T\cdot \mathcal P^{(L)}+M_R^T\cdot \mathcal P^{(R)}=0.
\end{equation}
Inside the WZW path integral 
we can then glue the $\phi^{L,R}$ theories together via a $\delta$-functional that enforces \eqref{WZWbcs}.  We can regard this $\delta$-functional as the limit of a sharply peaked Gaussian of the fields, which we can write, up to normalization, as
\begin{equation}
\prod_{\vec x\in\Sigma}\prod_I\delta[{M_L^T}^{IJ}\phi^{(L)}_J+{M_R^T}^{IJ}\phi^{(R)}_J]\sim\lim_{g_I\rightarrow \infty}\exp\left(-\sum_I\int_\Sigma d^2\vec x\,\frac{g_I}{2}\left({M_L^T}^{IJ}\phi^{(L)}_J+{M_R^T}^{IJ}\phi^{(R)}_J\right)^2\right).
\end{equation}
This introduces an effective quadratic interaction at infinite coupling into the Euclidean action.  Since this term is relevant we might loosely regard it as the IR end of the RG flow for a more generic gapping interaction.  In fact the specific form of the interaction is not important, only that it has a minimum that enforces \eqref{WZWbcs}.  For instance, we could have used a Sine-Gordon type interaction with a limit that $g_I\rightarrow\infty$:
\begin{equation}
\exp\left(-\sum_I\int_\Sigma d^2\vec x\,g_I\left(\cos\left(M_L^T\cdot\phi^{(L)}+M_R^T\cdot\phi^{(R)}\right)^I-1\right)\right)
\end{equation}
as introduced in \cite{Cano:2014pya}.

\section{Quantum Interfaces and Topological Entanglement Entropy}\label{QuantInt&TEE}
In the previous section, we considered the problem of finding appropriate interface conditions at a co-dimension one defect in $U(1)^N$ Chern-Simons theory. In this section, we want to consider the quantum version of this problem. In particular, we are interested in understanding the role of the above topological boundary conditions on the topological entanglement across the interface. Because we are interested in formulating calculations of entanglement in a gauge theory directly (i.e., we will not resort to surgery methods, as in \cite{Dong:2008ft}), we must confront the fact that the Hilbert space does not factorize spatially. To begin our discussion then, we explain how to deal with this by extending the Hilbert space.

\subsection{Extended Hilbert Space and Quantum Gluing}

Let us first consider the simplest possible case: we take a $U(1)$ Chern-Simons theory at level $k$ without defect, i.e., without a $\Sigma$ interface, and take space to be a 2-sphere. The entanglement cut then is as shown in Fig. \ref{fig:qg.pdf}(a), with $k$ the same on either side. That is, we wish to compute the entanglement entropy of the ground state under a partition $S^2 = D \cup_{S^1} \bar{D}$ of $S^2$ into two discs $D$ and $\bar{D}$. 
\myfig{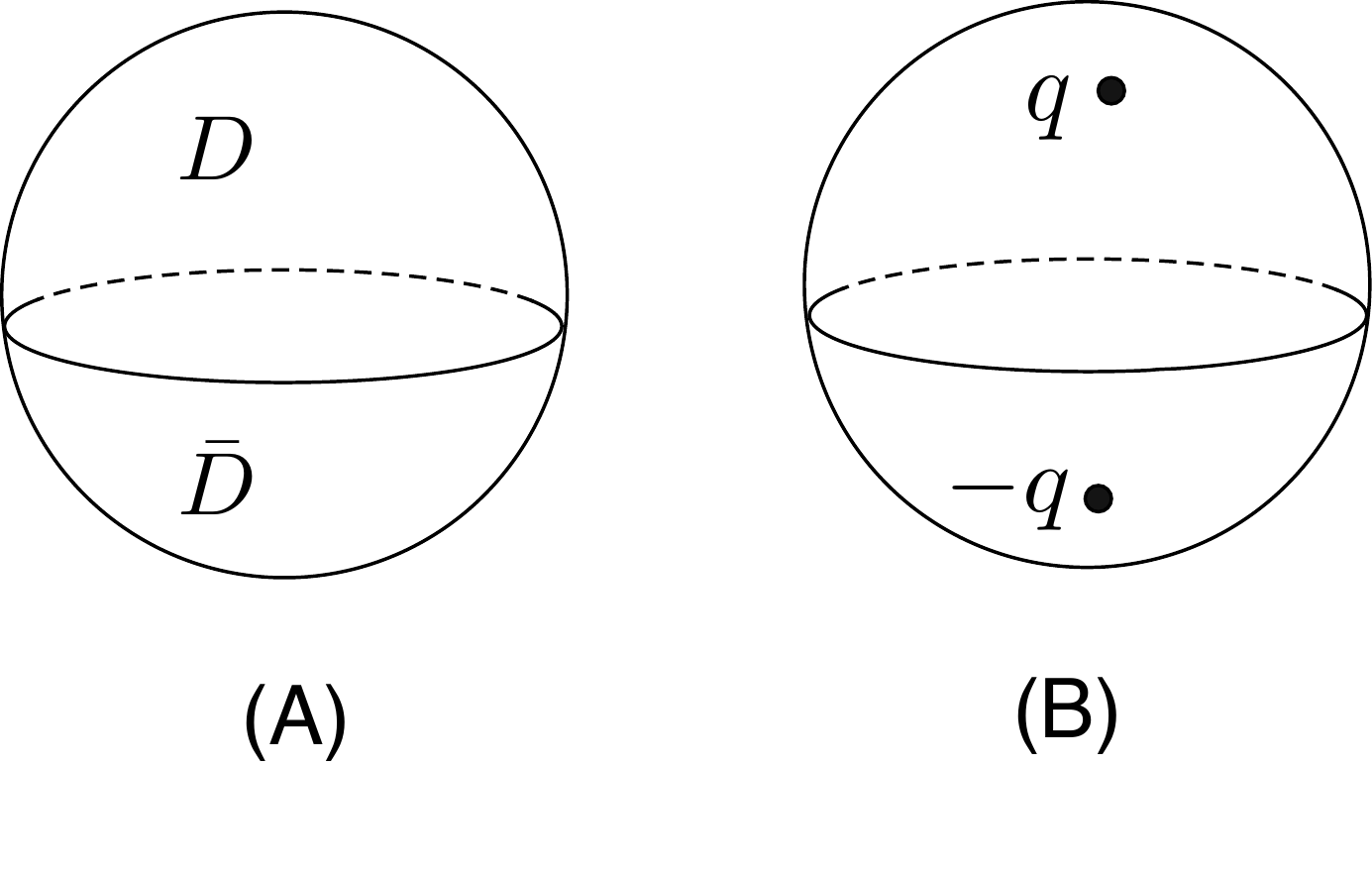}{6}{\textsf{(A) The spatial 2-sphere partitioned into two discs $D$ and $\bar D$. (B) The 2-sphere with an anyon $q$ in $D$ and $-q$ in $\bar D$.}}

Recall that the entanglement entropy is defined as follows: consider a Hilbert space $\mathcal{H}$ which admits a tensor factorization $\mathcal{H} = \mathcal{H}_{A} \otimes \mathcal{H}_{\bar{A}}$. Then given any unit-norm state $|\psi\rangle \in \mathcal{H}$, one constructs the reduced density matrix
\beq
\hat\rho_A = \mathrm{Tr}_{\mathcal{H}_{\bar A}} |\psi \rangle \langle \psi |.
\eeq
The entanglement entropy between $A$ and $\bar{A}$ is then defined as the von Neumann entropy of $\hat\rho_A$:
\beq
S_{EE}(A) =  -\mathrm{Tr}_{\mathcal{H}_{A}}\hat\rho_A\,\ln\,\hat\rho_A.
\eeq
Returning to our problem of computing the entanglement entropy between $D$ and $\bar{D}$, we immediately encounter a conceptual problem. For the entropy to be well-defined, we need
\beq \label{ni1}
\mathcal{H}_{S^2} \stackrel{?}{=} \mathcal{H}_{D} \otimes \mathcal{H}_{\bar{D}}.
\eeq
However, as was shown in \cite{Witten:1988hf}, the Hilbert space $\mathcal{H}_{S^2}$ of Chern-Simons theory on the 2-sphere (without Wilson lines) is one-dimensional, while the Hilbert space $\mathcal{H}_{D}$ on a disc (without Wilson lines) is a direct sum of Ka\v{c}-Moody modules corresponding to integrable representations of the $\mathfrak{u}(1)_k$ extended Ka\v{c}-Moody algebra 
(namely those labeled by $z\in \pi_1(U(1))=\mathbb Z$)\footnote{More precisely, $|z\neq0\rangle$ are desendents of the identity primary $|0\rangle$ via the action of the extended symmetry. Also, $\{J_n\}$ are generators of the $\mathfrak{u}(1)_k$ Ka\v{c}-Moody algebra which satisfy
$$
\left[J_m, J_n \right] =  \frac{k}{2} n\delta_{n+m,0}.
$$ 
For a more complete description of extended Ka\v c-Moody algebras, their commutation relations, and their representations, see \cite{Dong:2008ft, DiFrancesco:1997nk}.} \cite{Moore:1989yh, Elitzur:1989nr},
\beq
\mathcal H_D=\bigoplus_{z\in\mathbb Z}\mathcal H_D^{(z)}\qquad\qquad\mathcal{H}^{(z)}_D = \mathrm{span}\Big\{ |z\rangle,\; J_{-n} |z\rangle,\;J_{-n} J_{-m} |z\rangle \cdots \Big\}
\eeq
and in particular is infinite dimensional. Therefore, equation \eqref{ni1} is clearly not true. This problem is not new, but merely a manifestation of the standard problem with defining entanglement entropy in gauge theories -- the physical Hilbert space of gauge invariant states in gauge theories typically does not admit a simple tensor factorization. In this situation, one needs to choose a suitable redefinition of the entanglement entropy. A resolution to this problem that has appeared recently in the literature is to embed the gauge invariant Hilbert space into a larger Hilbert space that admits a tensor product factorization, $\mathcal H_{gauge\;inv}\subset \hat{\mathcal H}=\hat{\mathcal H}_A\otimes\hat{\mathcal H}_{\bar A}$.  The price one pays for this of course is that $\hat{\mathcal H}$ will contain states that are not gauge invariant. 

 We refer to this as the \emph{extended Hilbert space approach}.  While this approach has been detailed for gauge theories on the lattice, \cite{Donnelly:2011hn, Donnelly:2014gva}, and has been analogously been detailed in the continuum at the level of the classical phase space \cite{Donnelly:2016auv}, continuum examples are sparse (see, for instance, \cite{Donnelly:2014fua, Donnelly:2015hxa} for a path integral description in electromagnetism, and \cite{Donnelly:2016jet} for entanglement in closed string theory).  
Below we provide a clear and explicit example of implementing the extended Hilbert space approach in a continuum quantum gauge theory.  The basic idea is that even though equation \eqref{ni1} is false, it is nevertheless possible to realize the Hilbert space $\mathcal{H}_{S^2}$ as a subspace inside the extended Hilbert space $\mathcal{H}_{D} \otimes \mathcal{H}_{\bar{D}}$:
\beq
\mathcal{H}_{S^2}  \subset \mathcal{H}_{D} \otimes \mathcal{H}_{\bar{D}},
\eeq
where the precise injection depends on a choice of boundary conditions at the entanglement cut. To see how this works, let us consider the operator which generates infinitesimal gauge transformations on  $\mathcal{H}_{D} $:
\beq
\cQ_D(\lambda) =  \frac{k}{4\pi} \oint_{\pa D} \lambda \,A = \frac{k}{4\pi} \sum_n \lambda_n J_n,
\eeq
where $\lambda(\theta) = \sum_{n \in \mathbb{Z}} \lambda_n e^{in \theta}$ is the gauge transformation on $\pa D$, and $J_n$ are the generators of the $\mathfrak{u}(1)_k$ Ka\v c-Moody algebra. Note that the term gauge transformation here is a slight abuse of language because for Chern-Simons theory on the disc $D$, the ``gauge transformations'' at $\pa D$ are not really gauge symmetries, but are to be treated as global symmetries. That is, as per the discussion above, the Hilbert space furnishes a non-trivial representation of the generators of these transformations. 
Similarly, the generator of gauge transformations on $\mathcal{H}_{\bar D}$ is given by
\beq
\cQ_{\bar D}(\bar \lambda) =  \frac{k}{4\pi} \oint_{\pa \bar D} \bar\lambda \,A = \frac{k}{4\pi} \sum_n \bar\lambda_n \bar J_n.
\eeq
The gluing of the Chern-Simons theories on the two discs to form a Chern-Simons theory on the sphere involves a gluing map, which here can be specified by making an identification between $\{\lambda_n\}$ and $\{\bar\lambda_n\}$.
Let us consider the simplest such condition, $\lambda_n = \bar{\lambda}_{-n}$. Since gauge transformations are local, it is natural to define the generator for gauge transformations on the entire $S^2$ as the following operator on $\mathcal{H}_{D} \otimes \mathcal{H}_{\bar{D}}$
\beqn
\cQ_{S^2}(\lambda) &=& \cQ_D(\lambda)\otimes 1 + 1 \otimes \cQ_{\bar D}(\bar \lambda)\nonumber\\
&=& \frac{k}{4\pi} \sum_n\lambda_n \left( \,J_n\otimes 1 + 1\otimes \bar J_{-n}\right), 
\eeqn
where in the last equality above we have used the gluing condition. For a state $|\psi\rangle \in \mathcal{H}_{S^2}$ to be physical, we must impose the gauge invariance condition
\beq\label{ni2}
\cQ_{S^2}(\lambda) |\psi\rangle = 0,\;\;\; \Rightarrow \;\;\;\; \left( \,J_n\otimes 1 + 1\otimes \bar J_{-n}\right) |\psi\rangle = 0.
\eeq
Equation \eqref{ni2} can be regarded as the \emph{quantum gluing condition}.  Importantly, the quantum gluing condition uniquely identifies a one-dimensional subspace inside $\mathcal{H}_{D} \otimes \mathcal{H}_{\bar{D}}$, which is spanned by the \emph{Ishibashi state}\footnote{Often the Ishibashi condition is stated at the level of the Virasoro algebra, $(L_n\otimes 1-1\otimes \tilde L_{-n})|\psi\rangle=0$.  Here we have a refinement of this condition to the level of the current algebra.  Of course the conformal condition is additionally satisfied because the Virasoro generators will be given by a Sugawara construction: $L_n\sim \sum_{m}J_{m}J_{n-m}$ \cite{DiFrancesco:1997nk}.} corresponding to the identity operator\cite{Ishibashi:1988kg, Gaberdiel:2002my}.  We denote this state (and subsequent Ishibashi states) with the ``double bracket" notation:
\beq
|0\rrangle = \sum_{z\in \mathbb Z} \sum_{m} |z, m\rangle \otimes \overline{|z, m\rangle}. 
\eeq
where $m$ labels an orthonormal basis of states for the conformal module corresponding to the identity primary \cite{Ishibashi:1988kg}.  We have thus identified the physical Hilbert space $\mathcal{H}_{S^2}$ as a one-dimensional subspace of the extended Hilbert space $\mathcal{H}_{D} \otimes \mathcal{H}_{\bar{D}}$. As a consequence of this identification, we can now compute a well-defined  entanglement entropy between $D$ and $\bar{D}$ by tracing out $\mathcal{H}_{\bar{D}}$. This is essentially the computation of left-right entanglement entropy in Ishibashi states carried out in \cite{Das:2015oha, Wen:2016snr}. These papers showed by explicit computation that the left-right entanglement entropy in the Ishibashi state $| 0  \rrangle$ exactly reproduces the topological entanglement entropy of Chern-Simons theory on $S^2,$ where $S^2$ is bi-partitioned into two discs,
\begin{equation}
S_{ EE}(D)=-\frac{1}{2}\log k.
\end{equation}
While this result is well-known, the above formulation of the continuum extended Hilbert space definition of the entropy and the quantum gluing condition provides a universal explanation for why this calculation works, at the level of Chern-Simons theory, and is an important result of this article.\footnote{See also \cite{PhysRevLett.108.196402} for a different perspective involving gapping terms and quantum quench into a CFT.}

The above discussion can be generalized to the scenario where the state contains anyon insertions.  The charge of an anyon inserted in $D$ is only well-defined up to the image of the $K$-matrix and so for $U(1)_k$ we can choose its representative in $\mathbb Z_k$.  The corresponding Hilbert space, $\mathcal H_{D}(q)$ is the irreducible integral representation of the extended $\mathfrak{u}(1)_k$ Ka\v c-Moody algebra spanned by the chiral primary operator of charge $q$ and its descendants.  Correspondingly we must insert an anyon of charge $-q$ in $\bar{D}$ (see Fig. \ref{fig:qg.pdf}(B)). Here again, the Hilbert space on the 2-sphere $\mathcal{H}_{S^2}(q,-q)$ is one-dimensional, while the Hilbert space of the discs $\mathcal{H}_{D}(q)$ and $\mathcal H_{\bar D}(-q)$ are infinite dimensional. The gluing condition in this case is the same as before, and uniquely identifies $\mathcal{H}_{S^2}(q,-q) \subset \mathcal{H}_D(q) \otimes \mathcal{H}_{\bar D}(-q)$ as
\beq
| q  \rrangle  = \sum_{z\in\mathbb Z}\sum_{m} |z+q/k, m \rangle \otimes \overline{|z+q/k, m\rangle}. 
\eeq
The left-right entanglement entropy of this state once again matches the topological entropy of Chern-Simons theory in the presence of anyons \cite{Wen:2016snr}:
\begin{equation}
S_{ EE}(D, q)=-\frac{1}{2}\log k.
\end{equation}
(Strictly speaking, we are glossing over a subtlety in that the state $|q\rrangle$, as embedded in $\mathcal H_D\otimes \mathcal H_{\bar D}$, has infinite norm.  In order to properly normalize this state to unity, one must employ a suitable regularization.  This adds a non-universal divergent contribution to the entanglement \cite{Wen:2016snr}.) The generalization of these calculations to non-Abelian groups is entirely straightforward. 

\subsection{Interface entanglement entropy}

Now let us consider the more generic case where $D$ and $\bar D$ host $K$-matrix theories $K^{(L)}$ and $K^{(R)}$, respectively, and subject to the commensurability condition \eqref{injcond} for some primitive integral matrices $v^{(L)}$ and $v^{(R)}$. Additionally, we consider placing the entanglement cut right along the heterointerface. The insight in this case is that the Hilbert space on $S^2$, $\mathcal H_{S^2}[K^{(L)}, K^{(R)}]$ is still one-dimensional\footnote{Once gauge fields have been identified via topological boundary conditions, every Wilson loop operator on $S^2$ is contractible to the identity (see figure \ref{WLgluefig}).}, and so the primary task in defining the entanglement entropy across the interface is to identify the appropriate one-dimensional subspace within $\mathcal H_D\left[K^{(L)}\right]\otimes\mathcal H_{\bar D}\left[K^{(R)}\right]$.  The topological boundary conditions from Section \ref{ClassSympBCs} instruct us how to do so in the following way.

The tensor product space $\mathcal H_D\otimes\mathcal H_{\bar D}$ furnishes a representation of $U(1)^{2N}$ gauge transformations that do not vanish on $\pa D$.  From Section \ref{ClassSympBCs}, we know that the topological boundary conditions isolate an unbroken $U(1)^N\subset U(1)^{2N}$ at the interface\footnote{The injection of this unbroken group into $U(1)^{2N}$ is analogous to the identification of gauge tranformations on $D$ with gauge transformations on $\bar D$.} which have vanishing inner product with $\bs\Theta$ and so can be regarded as gauge transformations. This suggests that the appropriate \emph{quantum gluing condition} is to require 
$\mathcal H_{S^2}$ to be a gauge invariant subspace with respect to this unbroken $U(1)^N$.  To be specific, we write the generator of gauge transformations on $\mathcal H_D\left[K^{(L)}\right]$ for convenience as
\begin{equation}
\mathcal Q_D[\lambda^{(L)}_{I}]=\frac{K_{(L)}^{IJ}}{4\pi}\oint_{\pa D}\lambda^{(L)}_IA^{(L)}_J=\frac{K^{IJ}_{(L)}}{4\pi}\sum_{n}\lambda^{(L)}_{I;n}J^{(L)}_{J;n}
\end{equation}
and similarly on $\mathcal H_{\bar D}\left[K^{(R)}\right]$,
\begin{equation}
\mathcal Q_{\bar D}[\lambda^{(R)}_{I}]=\frac{K_{(R)}^{IJ}}{4\pi}\oint_{\pa \bar D}\lambda^{(R)}_{I}A^{(R)}_J=\frac{K_{IJ}^{(R)}}{4\pi}\sum_{n}\lambda^{(R)}_{I;n} J^{(R)}_{J;n}.
\end{equation}
Now we regard $\left({\lambda^{(L)}}^T,{\lambda^{(R)}}^T\right)^T$ as lying in the image of the injection 
$\mathbb P:\mathfrak t_0\hookrightarrow \mathfrak t_\Lambda$.  That is $\lambda^{(L)}_{n}=v^{(L)}\cdot\lambda_n$ and $\lambda^{(R)}_{n}=-v^{(R)}\cdot\lambda_{-n}$ for some $\lambda\in\mathfrak t_0$.  The generator of gauge transformations on $S^2$ can then be defined as
\begin{equation}
\mathcal Q_{S^2}(\lambda)=\frac{1}{4\pi}\sum_n\lambda_{I;n}\left[\left({M^{(L)}}^T\cdot J^{(L)}_{n}\right)^I\otimes 1-1\otimes \left({M^{(R)}}^T\cdot J^{(R)}_{-n}\right)^I\right].
\end{equation}
The state annihilated by $\mathcal Q_{S^2}(\lambda)$ for generic $\lambda_n$ then spans the gauge invariant Hilbert space on $S^2$.  This state can be regarded as the suitable $U(1)^N$ generalization of the Ishibashi state from the previous subsection.  In fact, defining $J^{(L)}_{n}\equiv v^{(L)}\cdot J_n$ and $ J^{(R)}_{n}\equiv -v^{(R)}\cdot \tilde J_n$, then the quantum gluing condition becomes
\begin{equation}
K_{eff}\cdot\left(J_n\otimes 1+1\otimes \tilde J_{-n}\right)\big|\psi\big\rangle=0.
\end{equation}
Thus, even in the inhomogeneous theory, the topological boundary conditions ensure that the entanglement across the interface is well-defined in terms of the left-right entanglement of the appropriate Ishibashi state. This left-right entanglement entropy can be straightforwardly computed, and one finds 
\beq
S_{EE}(D)  = -\frac{1}{2} \mathrm{log}\,|\mathrm{det}(K_{eff}) |,
\eeq
in agreement with the microscopic calculation of \cite{Cano:2014pya}. Of course, the definition of this Ishibashi state is intrinsically tied to the choice of boundary conditions, which is in turn directly manifested in the entropy.  Indeed, the calculation in  \cite{Cano:2014pya} involving gapping interactions in the coupled wire construction was in effect equivalent to computing the left-right entropy in the above Ishibashi state; note however that here, we have arrived at it from an entirely bulk Chern-Simons point of view. Again, we emphasize that the above discussion is only formal, because the Ishibashi states that we have identified are not normalizable and so an appropriate regularization needs to be employed. Therefore the entanglement entropy will have non-universal contributions that depend on the choice of regularization.  
In the following section we consider the computation from the perspective of the replica trick. In Section \ref{sectinhomoishi}, we will see that an analogous regularization naturally arises from a regulator surface enclosing the entanglement cut and supplementing the action by including \eqref{nontopBC} as a boundary term.

\section{Topological Entropy from the Replica trick} \label{sec4}
Given the above discussion, the role of TBCs in topological entanglement seems to be fairly straight-forward from the point of view of the left-right entanglement of the chiral edge theories.  Indeed, this is essentially the context of the calculation in \cite{Cano:2014pya}.  However,  it is instructive to illustrate the role of TBCs in alternative entanglement calculations involving the replica trick.  Let us briefly recall this method.  Given a reduced density matrix $\hat\rho$, the $n^{th}$ R\'enyi entropy is defined as
\begin{equation}
S^{(n)}=\frac{1}{1-n}\log\left(\mTr\ \hat\rho^n\right).
\end{equation}
The conventional von Neumann entanglement entropy is given by analytically continuing this to non-integer $n$ and taking the limit as $n\rightarrow 1$.  In the case where $\hat\rho$ is calculated from the identity sector with no anyon charges,
$S^{(n)}$ has the path integral representation
\begin{equation}
S^{(n)}=\frac{1}{1-n}\log\left(Z_n/Z_1^n\right)
\end{equation}
where $Z_n$ is the Euclidean path integral formed by cyclicly identifying replica fields.  This replica path integral typically possesses a conical singularity at the origin indicating an angular deficit that vanishes as $n\rightarrow1$.  Fortunately, Chern-Simons is a  topological theory --- \emph{even after gauge fixing} the path integral is metric-independent \cite{Blau:1989bq}.  Because of this we can choose a metric to smooth out the conical singularity and the entanglement entropy is given as the path integral on a (possibly complicated) replicated geometry.

In the simpler case without heterointerfaces, such replica path integrals were studied in \cite{Dong:2008ft} using systematic surgery methods for generic Chern-Simons theories along with generic choices of state, spatial topology, and entanglement cuts. 
The simplest scenario considered is the state on a spatial $S^2$, bipartitioned into hemispheres.  As shown in \cite{Dong:2008ft}, the replica geometry reproducing the $n^{th}$ R\'enyi entropy is topologically equivalent to $S^3$.  This path integral can be evaluated from modular properties of the theory via the following:  $S^3$ admits a Heegard splitting into two solid tori and to each torus the Chern-Simons path integral produces the identity state (since there are no Wilson lines inserted).  The path integral on $S^3$ can be interpreted as the overlap of identity states on separate tori with alternative cycles identified: $Z_{S^3}$ is the identity component of the modular $S$-matrix \cite{Witten:1988hf}:
\begin{equation}
Z_{S^3}={\mathcal S^0}_0.
\end{equation}
For $K$-matrix/Abelian topological theories, ${\mathcal S^0}_0=\left(\det K\right)^{-1/2},$ and in 
this simple case, the R\'enyi entropies are independent of $n.$ Hence, the entanglement entropy is given by the logarithm of ${\mathcal S^0}_0$. 

Direct path integral evaluations on the other hand are much more subtle: despite being a free Gaussian theory, these path integrals must be carefully gauge-fixed and the resulting determinants must be regularized.  For the rest of the paper we will be concerned with the theory defined on $S^3$.  For a homogeneous theory, the path integral is described carefully in Appendix \ref{appCShomoTEE}, reproducing the above result.

In the case of a theory with a heterointerface, the replica path integral is more complicated due to the proliferation of alternating topological phases.  As we will see, when considering the state on $S^2$, replica methods will again lead to a geometry diffeomorphic to $S^3$, but one which is ``striped" by alternating topological phases. We will now explore this replica geometry and discuss methods of computing the R\'enyi entropy.

\subsection{Interface Entanglement}\label{entcutalongint}
\begin{figure}[h!]
  \centering
  \begin{tabular}{ c c c }
    \includegraphics[width=.6\textwidth]{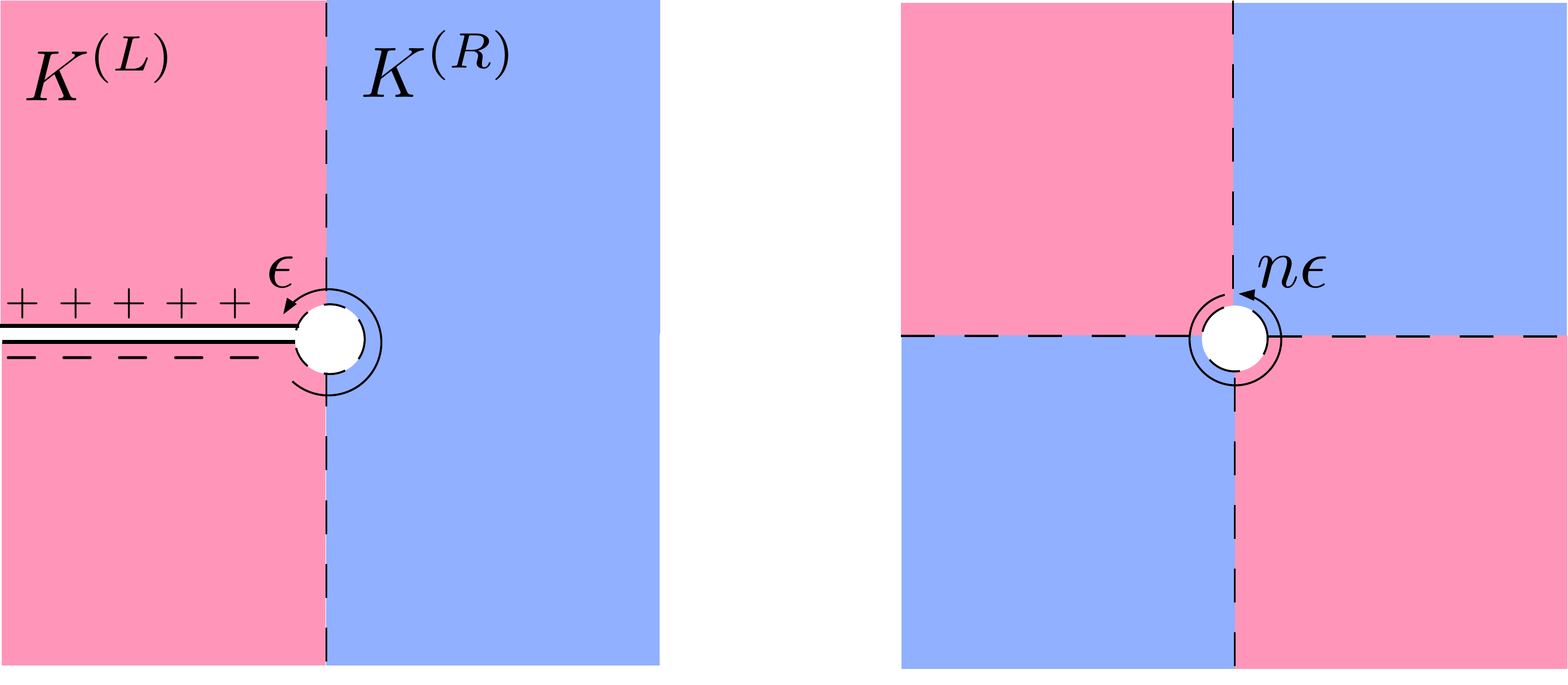} & \hspace{.8 cm} & \includegraphics[width=.21\textwidth]{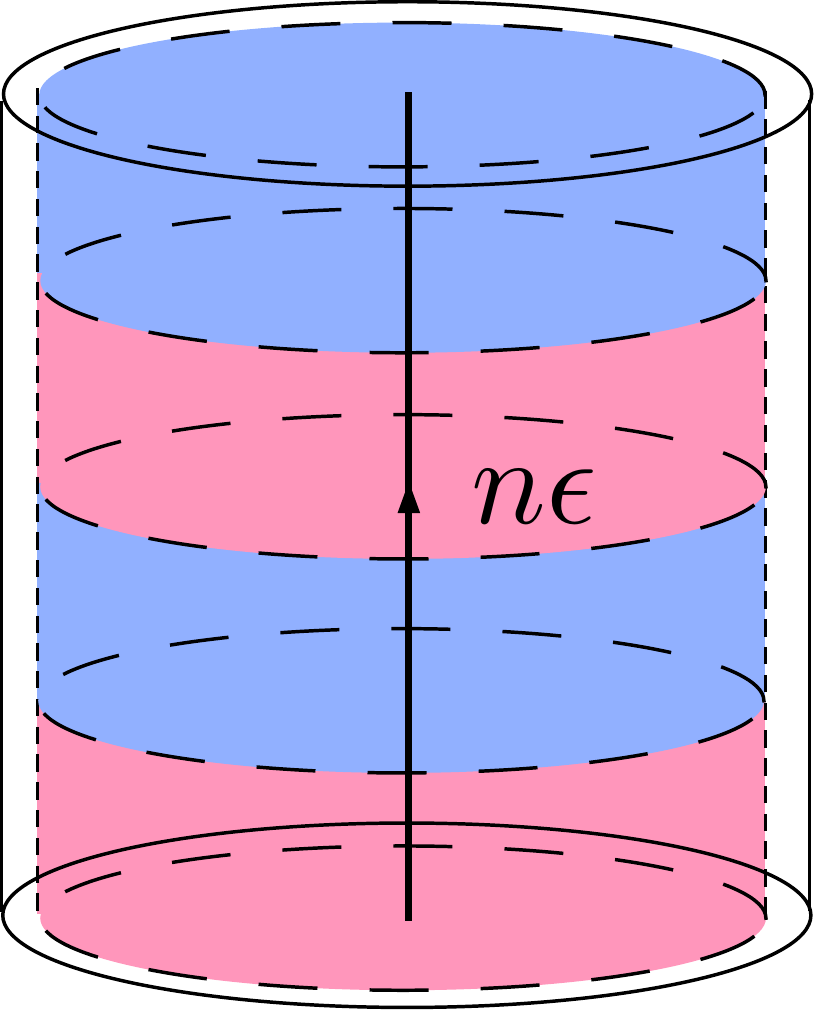}
    \end{tabular}
      \caption{\small{\textsf{(Left) A cartoon of the reduced density matrix after tracing out the $K^{(R)}$ phase. Regions with level matrix $K^{(L)}$ are denoted by pink shading in this and all other figures, and regions with level matrix $K^{(R)}$ are denoted by blue shading. The introduction of the regulator results in a ``keyhole" region in the reduced density matrix. (Middle) $\mTr(\hat\rho^n)$ is obtained by gluing $n$ copies of the first figure together cyclically. The figure represents this pictorially for $n=2$. In these illustrations, a transverse dimension, which can be interpreted as Euclidean time, 
      has been suppressed. (Right) This construction is conformally equivalent to a path integral on $S^1\times \mathbb H_2$, which can also be viewed as a solid torus. Top and bottom of this subfigure are identified. }}}
      \label{inhomoreplica}
\end{figure}

We regulate the replica trick calculations by excising a tubular neighbourhood of circumference $\epsilon$ about the entanglement cut. This results in a cutoff surface, as illustrated in Fig. \ref{inhomoreplica}. In this figure, we have decompactified $S^3$ to $\mathbb{R}^3$ and then suppressed a dimension for clarity. This excision has the effect of regulating the replica path integral for $tr\hat\rho^n$ by excising a cylinder with circumference $n\epsilon$.  This space is conformally equivalent\footnote{Explicitly, if we were to begin with the Euclidean metric $ds^2=d\tau^2+dr^2+r^2d\theta^2$ on $\mathbb{R}^3$, we can perform a Weyl transformation to a metric $ds^2=d\theta^2+\frac{d\tau^2+dr^2}{r^2}$. The regulator surface maps to the boundary of the Poincar\'e disc  $\mathbb H_2$.} to $S^1\times \mathbb H_2$ (a fact well utilized in the study of entanglement entropy.  See \cite{Casini:2011kv, Faulkner:2014jva, Hughes:2015ora}, for example).   
Equivalently, we can view this as a solid torus, as shown in Fig. \ref{inhomoreplica}(c), in which there are regions with alternating topological phases as we traverse the cycle of length $n\epsilon$.  

\begin{figure}[h!]
  \centering
    \includegraphics[width=\textwidth]{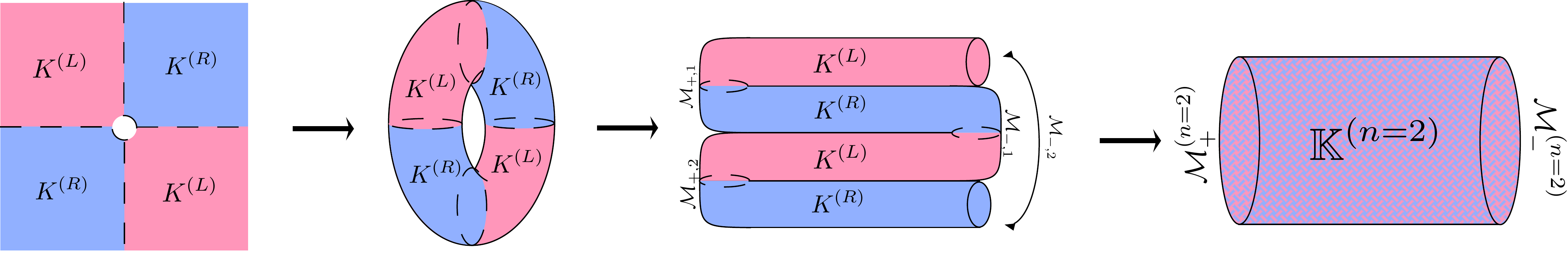}
      \caption{\small{\textsf{Replica path integral with $n=2$.  We first conformally map to $S_1\times \mathbb H_2$ and then ``fold" with parity transformations on the $K^{(R)}$ phases.  The edges at the fold determine topological boundary conditions ${\mathbb M}_{i,\pm}$.  Alternatively this can be packaged as a larger CS theory with a $2nN\times 2nN$ $K$-matrix, $\mathbb K^{(n)}$ and the edges have topological boundary conditions ${\mathbb M}_{\pm}^{(n)}$.}}}
      \label{inhomoreplicafolding}
\end{figure}

Conveniently, for the spatial bipartition of the state on $S^2$, this entire replica structure can also be encoded into a set of TBCs. 
To see this, we fold the replica theory 
with a parity transformation on each of the $K^{(R)}$ phases to achieve a $2nN$-component Chern-Simons theory with 
K-matrix
\begin{equation}
 \mathbb K^{(n)}=\left(\bigoplus_{q=1}^nK^{(L)}\right)\oplus\left(\bigoplus_{q=1}^n-K^{(R)}\right)=\left(\begin{array}{cccccccc}K^{(L)}&0&\ldots&0&0&0&\ldots&0\\
 0&K^{(L)}&\ldots&0&0&0&\ldots&0\\
 \vdots&\vdots&\ddots&\vdots&0&0&\ldots&0\\
 0&0&\ldots&K^{(L)}&0&0&\ldots&0\\
 0&0&\ldots&0&-K^{(R)}&0&\ldots&0\\
 0&0&\ldots&0&0&-K^{(R)}&\ldots&0\\
 0&0&\ldots&0&\vdots&\vdots&\ddots&\vdots\\
 0&0&\ldots&0&0&0&\ldots&-K^{(R)}\end{array}\right)
 \end{equation}
 on a cylinder of length $\epsilon$. For convenience, we have taken a basis in which $\mathbb{K}$ has each of the $K^{(L)}$ blocks side-by-side.  We denote the connection on this folded geometry as $\mathcal A$.  This is illustrated in Fig. \ref{inhomoreplicafolding}.

In this basis, each end of the cylinder, $\Sigma_\pm$, has topological boundary conditions denoted by matrices ${\mathbb M}_\pm^{(n)}$:
\begin{equation}
{\mathbb M}_+^{(n)}=\left(\begin{array}{cccccccc}K^{(L)}\cdot v^{(L)}&0&\ldots&0&K^{(R)}\cdot v^{(R)}&0&\ldots&0\\
0&K^{(L)}\cdot v^{(L)}&\ldots&0&0&K^{(R)}\cdot v^{(R)}&\ldots&0\\
\vdots&&\ddots&\vdots&\vdots&0&\ddots&\vdots\\
0&0&\ldots&K^{(L)}\cdot v^{(L)}&0&0&\ldots&K^{(R)}\cdot v^{(R)}\end{array}\right),\nonumber
\end{equation}
\begin{equation}\label{replicatopobcs}
{\mathbb M}_-^{(n)}=\left(\begin{array}{cccccccc}K^{(L)}\cdot v^{(L)}&0&\ldots&0&0&0&\ldots&K^{(R)}\cdot v^{(R)}\\
0&K^{(L)}\cdot v^{(L)}&\ldots&0&K^{(R)}\cdot v^{(R)}&0&\ldots&0\\
\vdots&\vdots&\ddots&\vdots&\vdots&\ddots&\vdots&\vdots\\
0&0&\ldots&K^{(L)}\cdot v^{(L)}&0&0&\ldots&0\end{array}\right)
\end{equation}
such that ${{\mathbb M}_\pm^{(n)}}^T\cdot\mathcal A$ pulls back to zero on $\Sigma_\pm$, respectively.  The replica path integral is then the path integral on this cylinder in the limit that $\epsilon$ goes to zero.

Let us introduce a coordinate $\tau$ transverse to $\Sigma_\pm$.  
Regarding $\mathcal A_\tau$ as a Lagrange multiplier, its path integral imposes the equation of motion 
as a constraint.  The path integral can then be written formally as
\begin{equation}\label{reducedreplicaPI}
Z_n=\!\!\!\int\!\!\mathcal\displaystyle\prod_{\cI=1}^{2nN}\prod_i\mathcal D\mathcal A_{i,\cI}\delta\left[\epsilon^{ij}\frac{\mathbb K^{(n)}}{4\pi}\cdot\pa_i\mathcal A_j\right]\delta_{\Sigma_\pm}\left[{\mathbb M}_\pm^{(n)}\cdot \mathfrak a^{\pm}_{i}\right]\exp\left(i\frac{\mathbb K^{(n),\cI\cJ}}{4\pi}\!\!\!\!\int \!\!\!\!d\tau d^2x\,\epsilon^{ij}\mathcal A_{i,\cI}\pa_\tau\mathcal A_{j,\cJ}\right)
\end{equation}
where by $\delta_{\Sigma_\pm}$ we actually mean a product of two delta functions enforcing the boundary conditions at each end and $\mathfrak a^{\pm}$ is the pullback of $\mathcal A$ to $\Sigma_\pm$, respectively.

 Eq. \eqref{reducedreplicaPI} can be readily rewritten as a path integral of pure gauge modes, an avenue that we will take shortly.  However let us see how much we can gain by working with the Chern-Simons field directly.
We write the index $\cI=(\{a_q\}_{q=1,...,n},\{\bar a_q\}_{q=1,...,n})$ where, for a fixed $q$, $a_q$ and $\bar a_q$ ranges from 1 to $N$.  This is to reflect the block structure of the $K$-matrix: $q$ indexes each replica and $a_q$ indexes a $K^{(L)}$ block within that replica 
 and similarly for $\bar a_q$ and the $-K^{(R)}$ blocks.  We then implement the following change of path integral variables: 
\begin{equation}
\mathcal A_{i,\cI}=
\left(\begin{array}{cc}\displaystyle\bigoplus_1^nv{(L)}&0\\0&\displaystyle\bigoplus_1^n\left(-v^{(R)}\right)\end{array}\right)\cdot\left(\begin{array}{c}\tilde A_{i,a_q}^{(L)}\\\tilde A_{i,\bar a_q}^{(R)}\end{array}\right).
\end{equation}
In doing so, there will be associated Jacobians in the measure -- 
determinants of $v^{(L)}$ and $-v^{(R)}$ raised to a regulator-dependent dimension, $n\mathscr P$.  Note that, in at least this na\"ive treatment, these powers of the Jacobians scale with $n$:
\begin{align}
Z_n=&\left(\det(v^{(L)})\det(-v^{(R)})\right)^{n\mathscr P}\!\!\!\int\displaystyle\prod_i\prod_{q=1}^n\prod_{a_q=1}^N\mathcal D\tilde A^{(L)}_{i,a_{q}}\prod_{\bar a_q=1}^N\mathcal D\tilde A^{(R)}_{i,\bar a_{q}}\left\{\delta\left[\epsilon^{ij}\frac{K_{eff}^{a_q b_q}}{4\pi}\cdot \pa_i\tilde A_{j,b_q}^{(L)}\right]\delta\left[\epsilon^{ij}\frac{-K_{eff}^{\bar a_q\bar b_q}}{4\pi}\cdot \pa_i\tilde A_{j,\bar b_{q}}^{(R)}\right]\right.\nonumber\\
&\times\left.\delta_{\Sigma_\pm}\left[B.C.s\right]\exp\left(i\sum_{q=1}^n\left(\frac{K_{eff}^{a_q b_q}}{4\pi}\int \epsilon^{ij}\tilde A^{(L)}_{i,a_q}\pa_\tau \tilde A^{(L)}_{j,b_q}-\frac{K_{eff}^{\bar a_q \bar b_q}}{4\pi}\int \epsilon^{ij}\tilde A^{(R)}_{i,\bar a_q}\pa_\tau \tilde A^{(R)}_{j,\bar b_q}\right)\right)\right\}
\end{align}
where we recall $K_{eff}={v^{(L)}}^T\cdot K^{(L)}\cdot v^{(L)}={v^{(R)}}^T\cdot K^{(R)}\cdot v^{(R)}$.  
Above we use the shorthand ``$\delta_{\Sigma_\pm}[B.C.s]$" to denote the delta functions enforcing the boundary conditions on $\Sigma_\pm$ in terms of the redefined fields.  

Let us elaborate on this result.  From the definitions of ${\mathbb M}_\pm^{(n)}$ we see that the boundary condition on $\Sigma_+$ enforces $K_{eff}\cdot\left(\mathfrak a^{+,(L)}_{i,a_q}-\mathfrak a^{+,(R)}_{i,\bar a_q}\right)=0$ for each $q=1,\ldots, n,$ while on $\Sigma_-$ we have a similar condition but cyclicly permuted: $K_{eff}.\left(\mathfrak a^{-,(L)}_{i,a_q}-\mathfrak a^{-,(R)}_{i,\bar a_{q-1}}\right)=0$, where we use the shorthand $\bar a_0\equiv\bar a_n$.  Then up to another Jacobian (with a power once again scaling with $n$), the $\delta$-functionals enforce that $\tilde A^{(L)}$ and $\tilde A^{(R)}$, together, become a continuous field: passing through $\Sigma_+$, $\tilde A^{(L)}_{a_1}$ transitions into $\tilde A^{(R)}_{\bar a_1},$ which when passing through $\Sigma_-$ moves to $\tilde A^{(L)}_{a_2}$ and so on, cyclicly until $\tilde A^{(R)}_{\bar a_n}$ transitions back to $\tilde A^{(L)}_{a_1}$.  We now recognize that the action and flatness constraints of these fields are that of a single homogeneous theory with $K$-matrix $K_{eff}$ on the $n$-replicated manifold:
\begin{equation}
Z_n=\left(\det(v^{(L)})\det(-v^{(R)})\right)^{n\mathscr P}\left(\det K_{eff}\right)^{-n\mathscr P'}Z_n[K_{eff}].
\end{equation}
From here it is straightforward to see that in calculating the R\'enyi entropy all that matters is the \emph{effective homogeneous theory} for $K_{eff}$, because the other determinants drop out:
\begin{equation}
S^{(n)}=\frac1{1-n}\log\left(\frac{Z_n}{Z_1^n}\right)=\frac1{1-n}\log\left(\frac{Z_n[K_{eff}]}{Z_1[K_{eff}]^n}\right).
\end{equation}
Having thus `homogenized' the theory, we 
take the geometric regulator to zero
and thus 
find that the topological entanglement entropy across the heterointerface is
\begin{equation}
S_{EE}=S^{(n)}=-\frac{1}{2}\log\left|\det K_{eff}\right|.
\end{equation}

\subsection{Wess-Zumino-Witten description of the replica path integral}\label{sectinhomoishi}

The above calculation should be regarded as a formal result.  In particular, 
we did not carefully specify a regularization scheme when manipulating functional determinants, noting only the important fact of their dependence on $n$.  To refine this, let us go back and evaluate \eqref{reducedreplicaPI} as a path integral of WZW fields living on the regulator surface. 
In doing so, we will show that the replica calculation reduces to a transition amplitude between conformal boundary states. Indeed, our use of TBCs determines these boundary states to be the familiar Ishibashi states of the replicated theory.  Given the discussion in Section \ref{QuantInt&TEE}, this bears resemblance to the familiar LREE computations.  However, we emphasize that this is not, {\it a priori}, a boundary LREE calculation, but instead a precise rewriting of the CS path integral as a CFT transition amplitude.  We then regard the following as a complementary physical picture to the discussion in Section \ref{QuantInt&TEE}.


In the previous section, we 
packaged the replica path integral into a single theory on a solid cylinder with a $U(1)^{2nN}$ gauge field $\mathcal A$ and a $2nN\times 2nN$ K matrix, $\mathbb K^{(n)}$.  In doing so, TBCs given by ${\mathbb M}_\pm^{(n)}$ 
were defined on the interface surfaces $\Sigma_\pm$ separated by Euclidean time $\epsilon$:  
\begin{equation}\label{reducedreplicaPI2}
Z^{\mathbb R^3}_n[K^{(L)}, K^{(R)}, M^{(L)}, M^{(R)}]=\lim_{\epsilon\rightarrow 0}Z^{[0,\epsilon]\times\mathbb H_2}[\mathbb K^{(n)},{\mathbb M}^{(n)}_{\Sigma_\pm},\epsilon].
\end{equation}
The flatness constraint \eqref{reducedreplicaPI} is solved locally by writing 
\beq
\mathcal A_{i,\cI}=\pa_i\phi_\cI.
\eeq 
The action of these pure gauge modes is localized on the regulating surface, $\mathcal R=[0,\epsilon]\times \pa\mathbb H_2$.  As discussed at the end of Section \ref{ClassSympBCs}, $\phi_\cI$ need not be single-valued.  Since $\pa\Sigma_{\pm}\simeq S^1$ the bosons possess winding periodicities labelled by a $2nN$ integer vector $\mathcal P^{(n)}$.  Shortly we will see that in order to give dynamics to the theory on $\mathcal R$, we can introduce the non-topological term $\frac{\mathbb V^{\cI\cJ}}{4\pi}\int_{\mathcal R}a_{\cI}\wedge \ast_{\mathcal R}a_{\cJ}$, where $\ast_{\mathcal R}$ denotes the Hodge star for a Riemannian metric on $\mathcal R$. 
To be definite, let the coordinate on $\pa\mathbb H_2$ be $\sigma$ ranging from $0$ to $\ell$.  We then choose the metric to be orthonormal with respect to $\pa_\tau$ and $\pa_\sigma$.  The limit in \eqref{reducedreplicaPI2} can then be expanded as a series in $\epsilon/\ell\rightarrow 0$.  We take the action on $\mathcal R$ to contain the terms
\begin{equation}\label{ActionReplicaEdge}
S=\frac{{\mathbb K^{(n)}}^{\cI\cJ}}{4\pi}\int_{\mathcal R} d\tau\,d\sigma\; \pa_\tau\phi_\cI\pa_\sigma\phi_\cJ-i\frac{\mathbb V^{\cI\cJ}}{4\pi}\int_{\mathcal R} d\tau\,d\sigma\; \pa_\sigma\phi_\cI\pa_\sigma\phi_\cJ.
\end{equation}
Note that the first term in \eqref{ActionReplicaEdge} is first order in either derivative; in quantizing along either the interval $[0,\epsilon]$, or along $\pa\mathbb H_2$ the Hamiltonian will exactly vanish.  Thus we see that $\mathbb V^{\cI\cJ}$ supplements this theory with a Hamiltonian.  However, the choice of $\mathbb V$ is non-universal and not expected to affect the outcome of the topological entanglement.  Later we will choose it for convenience.

Let us now map out how we will proceed.  We will shortly show that the problem at hand is equivalent to the partition function of a 2d CFT at central charge $c+\tilde c=2nN$ on a finite cylinder.  In the context of boundary CFTs \cite{Cardy:1989ir, Cardy:2004hm}, it is a standard result  that this partition function can either be viewed as the trace of a state defined along the interval $[0,\epsilon],$ or as the Euclidean transition amplitude from a conformal boundary state on $\pa\Sigma_-$ to a conformal boundary state on $\pa\Sigma_+$.  It is most convenient to formulate our problem in the latter language.  That is we will express \eqref{reducedreplicaPI2} as  
\begin{equation}
Z_n=\lim_{\epsilon/\ell\rightarrow0}\llangle\pa\Sigma^{(n)}_-|e^{-\frac{2\pi\epsilon}{\ell}\left(L^{(n)}_0+\tilde L^{(n)}_0-\frac{nN}{12}\right)}|\pa\Sigma^{(n)}_+\rrangle
\end{equation}
where $|\pa\Sigma^{(n)}_\pm\rrangle$ are suitable boundary states  and  $L^{(n)}_0$ and $\tilde L^{(n)}_0$ are the Virasoro generators.  This approach will allow us to evaluate $Z_n$ in canonical quantization 
and thus avoid the subtlety of functional Jacobians coming from field redefinitions.

Informed from our discussion in the previous section, we make the field redefinition
\begin{equation}\label{ishifieldredef}
\phi=\left(\begin{array}{cc}\displaystyle\bigoplus_1^nv^{(L)}&0\\0&\displaystyle\bigoplus_1^n\left(-v^{(R)}\right)\end{array}\right)\cdot\phi'\equiv\mathfrak v\cdot\phi'
\end{equation}
for which the action simplifies to
\begin{equation}
S=\frac{{\mathbb K_{eff}^{(n)}}^{\cI\cJ}}{4\pi}\int d\tau\,d\sigma\; \pa_\tau\phi'_\cI\pa_\sigma\phi'_\cJ-i\frac{\mathbb V'^{\cI\cJ}}{4\pi}\int d\tau\,d\sigma\; \pa_\sigma\phi'_\cI\pa_\sigma\phi'_\cJ
\end{equation}
for $\mathbb K_{eff}^{(n)}\equiv \left(\displaystyle\bigoplus_1^n K_{eff}\right)\oplus\left(\displaystyle\bigoplus_1^n -K_{eff}\right)$ and $\mathbb V'$ such that $\mathbb V=\mathfrak v^T\cdot\mathbb V'\cdot \mathfrak v$. 
We choose a frame, $E$, (and coframe, $F$) for the $K$-matrix, defined by
\begin{equation}\label{Kframe}
{\mathbb K_{eff}^{(n)}}^{\cI\cJ}={E^{(n)}}^\cI_A\eta^{AB}{E^{(n)}}^\cJ_B\qquad\qquad\qquad\qquad {F^{(n)}}^A_\cJ{E^{(n)}}^\cI_A=\delta^\cI_\cJ
\end{equation}
where $\eta$ is a signature $(nN,nN)$ diagonal matrix of $\pm 1$.  We then define fields $\Phi_A={E^{(n)}}^\cI_A{\phi'}_\cI$ and velocity, $v^{AB}={F^{(n)}}^A_I\mathbb {V'}^{\cI\cJ} {F^{(n)}}^B_\cJ$ in this frame.  Because $v^{AB}$ is non-universal, we will simply choose it for convenience to be proportional to $\delta^{AB}$.

This system is easily quantized. It will be convenient to split the index $A$ in an way analogous to the analysis in Section \ref{entcutalongint}.  That is, we take $A=(\{a_q\}_{q=1\ldots n}, \{\bar a_q\}_{q=1\ldots n})$ with $a_q$ and $\bar a_q$ for a given $q$ 
ranging from 1 to $N$.  The mode expansion for $\Phi_A$ is then
\begin{equation}\label{JFmodeexp}
\Phi_A(\tau,\sigma)=\varphi_A(\tau)+\frac{i2\pi}{\ell}\left(\mathcal P_{L,A}\left(\tau-i\sigma\right)-\mathcal P_{R,A}\left(\tau+i\sigma\right)\right)+i\!\!\!\!\!\!\!\!\sum_{k=-\infty,k\neq0}^{\infty}\frac{1}{n}\left(\alpha^{(L)}_{A,k}\,e^{\frac{2\pi k}{\ell}(\tau-i\sigma)}+\tilde{\alpha}^{(R)}_{A,k}\,e^{\frac{2\pi k}{\ell}(\tau+i\sigma)}\right)
\end{equation}
where $\mathcal P_{L,A}$ and $\alpha_{A,k}^{(L)}$ are non-zero only in the upper block, and vice-versa for $\mathcal P_{R,A}$ and $\tilde\alpha^{(R)}_{A,k}$.  That is,  $\mathcal P_{L,A}=\delta_A^{a_q}\mathcal P_{L,a_q}$, $\mathcal P_{R,A}=\delta_A^{\bar a_q}\mathcal P_{R,\bar a_q}$, $\alpha^{(L)}_{A,k}=\delta_A^{a_q}\alpha^{(L)}_{a_q,k}$, and $\tilde\alpha^{(R)}_{A,k}=\delta_A^{\bar a_q}\tilde\alpha^{(R)}_{\bar a_q,k}$.  
Modes within the same replica block obey the commutation relations
\begin{equation}
[\varphi_A(0),\mathcal P_B]=i\eta_{AB}\qquad [\alpha^{(L)}_{a_{q_1},m},\alpha^{(L)}_{b_{q_2},k}]=m\,\delta_{q_1,q_2}\delta_{a_{q_1}b_{q_2}}\,\delta_{m+k}\qquad[\tilde\alpha^{(R)}_{\bar a_{q_1},m},\tilde\alpha^{(R)}_{\bar b_{q_2},k}]=m\,\delta_{q_1,q_2}\delta_{\bar a_{q_1}\bar b_{q_2}}\delta_{m+k}.
\end{equation}
The normal-ordered Hamiltonian of the system is
\beq
H=\frac{1}{4\pi}\int_0^\ell d\sigma :\pa_\sigma\hat\Phi_A\delta^{AB}\pa_\sigma\hat\Phi_B:=\frac{2\pi}{\ell}\left(L^{(n)}_0+\tilde L^{(n)}_0-\frac{nN}{12}\right).
\eeq
where
\begin{equation}
L_0^{(n)}
=\sum_{q=1}^n\sum_{a_q,b_q=1}^{N}\delta^{a_q, b_q}\left(\frac{1}{2}{\hat{\mathcal P}_{L,a_q}^{(n)}}\hat{\mathcal P}_{L,b_q}^{(n)}+\sum_{k=1}^\infty\alpha^{(L)}_{a_q,-k}\alpha^{(L)}_{a_q,k}\right)
\end{equation}
\begin{equation}
\tilde L_0^{(n)}
=\sum_{q=1}^n\sum_{\bar a_q,\bar b_q=1}^{N}\delta^{\bar a_q, \bar b_q}\left(\frac{1}{2}{\hat{\mathcal P}_{R,\bar a_q}^{(n)}}\hat{\mathcal P}_{R,\bar b_q}^{(n)}+\sum_{k=1}^\infty\tilde\alpha^{(R)}_{\bar a_q,-k}\tilde\alpha^{(R)}_{\bar a_q,k}\right)
\end{equation}
We specify the boundary states, $|\pa\Sigma_\pm\rrangle$ by the constraints that they satisfy the replica TBCs
\begin{equation}
\left({\mathbb M^{(n)}}^T_+\cdot\pa_\sigma\hat\phi\right)|\pa\Sigma_+\rrangle=\left({\mathbb M^{(n)}}^T_-\cdot\pa_\sigma\hat\phi\right)|\pa\Sigma_-\rrangle=0.
\end{equation}
These states will be labeled by their zero mode eigenvalues, $\{\mathcal P_\pm^{(n)}\}$, which, in the current frame, are respectively spanned by a collection of integer vectors 
 $\{z_\pm^i\}_{i=1,2,\ldots,n}\in\mathbb Z^N$:
\begin{equation}\label{restlattvecs}
\mathcal P^{(n)}_+=\left(\begin{array}{c}\mathcal P^{(n)}_{L,+}\\\mathcal P^{(n)}_{R,+}\end{array}\right)=\left(\begin{array}{c} z^1_+\\z^2_+\\\vdots\\ z^n_+\\ z^1_+\\z^2_+\\\vdots\\z^{n}_+\end{array}\right)\qquad\qquad\qquad \mathcal P^{(n)}_-=\left(\begin{array}{c}\mathcal P^{(n)}_{L,-}\\\mathcal P^{(n)}_{R,-}\end{array}\right)=\left(\begin{array}{c}z^n_-\\ z^1_-\\\vdots\\z^{n-1}_-\\z_-^1\\z^{2}_-\\\vdots\\z^{n}_-\end{array}\right).
\end{equation}
Additionally, the oscillator portion of the boundary states are subject to the boundary condition 
\begin{equation}\label{replicaishibcs}
\left(\alpha^{(L)}_{a_q,k}-\tilde\alpha^{(R)}_{\bar a_q,-k}\right)|\pa\Sigma_+\rrangle=0\qquad\qquad \left(\alpha^{(L)}_{a_q,k}-\tilde \alpha^{(R)}_{\bar a_{q-1},-k}\right)|\pa\Sigma_-\rrangle=0
\end{equation}
for each $q=1,\ldots, n$.  In the second equation in eq. (\ref{replicaishibcs}), one should regard $\bar a_0\equiv \bar a_n$. These boundary conditions are reminiscent of 
those in Section \ref{QuantInt&TEE}, replicated and cyclicly identified,\footnote{Indeed, on $\mathcal R$, 
it is the oscillator modes of $\pa_\sigma\phi$ that span the current algebra: $J_k\sim \alpha_k$, $\tilde J_k\sim \tilde \alpha_k$.} and so we see the familiar role of the Ishibashi state appearing, this time from the replica trick.  The full boundary state at $\Sigma_+$ can be written as the superposition over all occupations of Fock states with equal left- and right-occupation:
\begin{align}
|\pa\Sigma_{+}\rrangle=&\!\!\!\!\sum_{\{z^i_+\}\in \mathbb Z^{N}}\sum_{\{m_{a_qk}\}=1}^\infty\prod_{q=1}^n\prod_{a_q=\bar a_q=1}^{N}\prod_{k=1}^{\infty}\frac{1}{m_{a_qk}!}\left(\frac{1}{k}\alpha_{a_q,-k}^{(L)}\tilde\alpha_{\bar a_q,-k}^{(R)}\right)^{m_{a_qk}}|\{\mathcal P^{(n)}_+(z^i_+)\}\rangle\nonumber\\
\equiv&\!\!\!\!\sum_{\{z^i_+\}\in \mathbb Z^{N}}\sum_{\{m_{a_qk}\}=1}^\infty|\{\mathcal P_+^{(n)}(z^i_+)\};\vec m_{a_qk}\rangle
\end{align}
and similarly for $\pa\Sigma_-$
\begin{align}
|\pa\Sigma_{-}\rrangle=&\!\!\!\!\sum_{\{z^i_-\}\in \mathbb Z^{N}}\sum_{\{m_{a_qk}\}=1}^\infty\prod_{q=1}^n\prod_{a_q=\bar a_q=1}^{N}\prod_{k=1}^{\infty}\frac{1}{m_{a_qk}!}\left(\frac{1}{k}\alpha_{a_q,-k}^{(L)}\tilde\alpha_{\bar a_{q-1},-k}^{(R)}\right)^{m_{a_qk}}|\{\mathcal P^{(n)}_-(z^i_-)\}\rangle\nonumber\\
\equiv&\!\!\!\!\sum_{\{z^i_-\}\in \mathbb Z^{N}}\sum_{\{m_{a_qk}\}=1}^\infty |\{\mathcal P_-^{(n)}(z^i_-)\};\vec m_{a_qk}\rangle.
\end{align}
The Euclidean evolution of $|\pa\Sigma_+\rrangle$ to the other end of the interval gives us
\begin{align}
e^{-\frac{2\pi\epsilon}{\ell}\left(L_0^{(n)}+\tilde L_0^{(n)}-\frac{nN}{12}\right)}&|\pa\Sigma^{(n)}_+\rrangle\nonumber\\
&=e^{\frac{2\pi\epsilon}{\ell}\frac{nN}{12}}\!\!\!\!\sum_{\{z^i\}\in\mathbb Z^N}\!\!\!\!e^{-\frac{2\pi\epsilon}{\ell}\left(\sum_{i=1}^n{z^i_+}^T\cdot K_{eff}\cdot z^i_+\right)}\prod_{q=1}^n\prod_{a_q=1}^{N}\prod_{k=1}^\infty\sum_{m_{a_qk}=1}^\infty e^{-\frac{4\pi\epsilon}{\ell}k\,m_{a_qk}}|\{\mathcal P_+^{(n)}(z^i_+)\};\vec m_{a_qk}\rangle.
\end{align}
Now it is easy to see what will happen in the inner product with $\llangle\pa\Sigma_-|$. In the zero mode sector, the inner product on the lower block (i.e., $\mathcal P_R^{(n)}$) will enforce $z_+^i=z_-^i$ while the upper ($\mathcal P_R^{(n)}$) block will enforce $z_+^i=z_-^{i-1}$.  The sum over the integer vectors will then collapse to a sum over a single integer vector $z$.  Similarly because the oscillators of $|\pa\Sigma_-\rrangle$ have been cyclicly permuted from the definition of $|\pa\Sigma_+\rrangle$, the only nonzero portion of this inner product comes from the occupations satisfying $m_{a_1k}=m_{a_2k}=\ldots=m_{a_nk}\equiv m_{ak}$ and so this product also collapses:
\begin{equation}
\llangle \pa\Sigma_-^{(n)}|e^{-\frac{2\pi\epsilon}{\ell}\left(L_0^{(n)}+\tilde L_0^{(n)}-\frac{nN}{12}\right)}|\pa\Sigma_+^{(n)}\rrangle=e^{\frac{4\pi\epsilon}{\ell}\frac{nN}{24}}\sum_{z\in\mathbb Z^N}e^{-\frac{2\pi\epsilon}{\ell}\left(n\,z^T\cdot K_{eff}\cdot z\right)}\prod_{a=1}^{N}\prod_{k=1}^\infty\sum_{m_{ak}=1}^\infty e^{-\frac{4\pi\epsilon}{\ell}n\,k\,m_{ak}}.
\end{equation}
As is familiar, the sum over occupations, $m_{ak}$, along with the oscillator products, and the overall coefficient, gives the Dedekind $\eta$-function, $\left(\eta\left(i\frac{2n\epsilon}{\ell}\right)\right)^{-N}$ \cite{DiFrancesco:1997nk}, while the sum over $z$ gives the Riemann $\theta$-function associated to the matrix $K_{eff}$ \cite{mumford1983tata}.  We then have that the replica path integral becomes
\begin{equation}
Z_n=\lim_{\epsilon/\ell\rightarrow 0}\left(\eta\left(i\frac{2n\epsilon}{\ell}\right)\right)^{-N}\vartheta_{(K_{eff})}\left(0\Big|i\frac{2n\epsilon}{\ell}\right).
\end{equation}
Dividing this by the normalization
\begin{equation}
\frac{Z_n}{Z_1^n}=\lim_{\epsilon/\ell\rightarrow 0}\left(\eta\left(i\frac{2n\epsilon}{\ell}\right)\right)^{-N}\left(\eta\left(i\frac{2\epsilon}{\ell}\right)\right)^{nN}\vartheta_{(K_{eff})}\left(0\Big|i\frac{2n\epsilon}{\ell}\right)\vartheta^{-n}_{(K_{eff})}\left(0\Big|i\frac{2\epsilon}{\ell}\right).
\end{equation}
Using the modular properties, $\eta(\tau)=\left(-i\tau\right)^{-1/2}\eta(-1/\tau)$ and $\vartheta_\Omega(0|\tau)={\det}^{-1/2}\left(-i\tau\Omega\right)\vartheta_\Omega(0|-1/\tau),$ we can expand this in the limit that $\epsilon/\ell\rightarrow0$:
\begin{align}
\frac{Z_n}{Z_1^n}&=\lim_{\epsilon/\ell\rightarrow0}\text{det}^{-1/2}\left(K_{eff}\right)\text{det}^{n/2}\left(K_{eff}\right)\left(\eta\left(i\frac{\ell}{2n\epsilon}\right)\right)^{-N}\!\!\!\!\left(\eta\left(i\frac{\ell}{\epsilon}\right)\right)^{nN}\!\!\!\!\vartheta_{(K_{eff})}\left(0\Big|i\frac{\ell}{n\epsilon}\right)\vartheta^{-n}_{(K_{eff})}\left(0\Big|i\frac{\ell}{\epsilon}\right)\nonumber\\
&=\lim_{\epsilon/\ell\rightarrow 0}e^{\frac{\pi N\ell}{24\epsilon}\frac{(1-n)(1+n)}{n}}\text{det}^{(n-1)/2}(K_{eff}).
\end{align}
The R\'enyi entropy is given as
\begin{equation}
S^{(n)}=\frac{1}{1-n}\log\left(\frac{Z_n}{Z_1^n}\right)=\lim_{\epsilon/\ell\rightarrow0}\left\{\frac{\pi N}{24}\frac{1+n}{n}\frac{\ell}{\epsilon}-\frac{1}{2}\log\left|\text{det}(K_{eff})\right|\right\}.
\end{equation}
and the entanglement entropy is the $n\rightarrow1$ limit of this:
\begin{equation}
S_{EE}=\lim_{\epsilon/\ell\rightarrow0}\left\{\frac{\pi N}{12}\frac{\ell}{\epsilon}-\frac{1}{2}\log\left|\text{det}(K_{eff})\right|\right\}.
\end{equation}
We see that the piece independent of the cutoff is precisely the determinant of the effective $K$ matrix. This concludes our assertion that not only do TBCs modify the topological entanglement, but in precisely the same fashion as in \cite{Cano:2014pya}.

\section{Discussion}

In this paper we have addressed the gapped interfaces between topological phases from a bulk Chern-Simons approach.  Central to this discussion is the existence of topological interface conditions.  The algebraic properties of these boundary conditions are closely related to the existence of the gapping potentials in the chiral boundary theory. 
This is a restatement of the fact that the ``glue-ability" of two Chern-Simons theories is equivalent to ``gappability" of the interface chiral modes.  Although derived from a classical symplectic analysis, we showed that TBCs lead to a natural quantum criteria for isolating the Chern-Simons ground state in the extended Hilbert space approach to entanglement.  From there we showed that the signatures of the TBCs are seen in the entanglement entropy across a heterointerface.  In particular, the TBCs can be thought of as identifying which linear combinations of gauge fields can permeate the interface.  The effective theory of these gauge fields at the boundary is characterized by a new $K$-matrix that we call $K_{eff}$ and the topological entanglement probes this matrix.  These results nicely corroborate with known results in the condensed matter literature.  

There are several natural extensions to the program that we have initiated here.  First, we have focused on states of Chern-Simons theory defined on 
constant time slices having the topology of $S^2$.  Although, this is sufficient for illustrating the sensitivity of the TEE to the interface conditions, this is a drastic simplification to the wealth of states we could construct in Chern-Simons theory.  In particular we can define a state on any Riemann surface.  Even in a homogeneous theory, the entanglement structure on such surfaces is more interesting: the Hilbert space is no longer one-dimensional and so there are multiple sectors from which one can define an entanglement entropy, and additionally there may be more than one topologically inequivalent way of bi-partitioning the surface.  The analysis of heterointerfaces adds additional, interesting structures to this problem: in particular, if a bi-partition requires the interface to consist of multiple components, one can imagine choosing different TBCs on each component.  We expect the reasoning in Section \ref{QuantInt&TEE} to play a guiding role in such analysis; indeed one may even hope to develop a set of rules akin to ``surgery for hetero-interfaces."  

A second generalization is to explore the gluing of two non-Abelian topological phases.  
Although in this context there is no natural notion of a $K$-matrix, only an integer $k$, the classification of TBCs should not be discarded as simple.  We again expect isolating a half-dimensional unbroken gauge symmetry to play a central role in this analysis; e.g., if joining phases with groups, $G$ and $\tilde G$, the TBCs should define a half-dimensional Lie group $H$ immersed into $G\times \tilde G$, while also satisfying algebraic properties involving the levels.  This, itself, presents an interesting geometric problem.  For discussions of entanglement, Wilson line contributions to the entropy provide an additional subtlety to this problem 
 that does not arise in the Abelian context.  

\vskip .5cm

{\bf Acknowledgements:} 
We would like to acknowledge helpful conversations with Vijay Balasubramanian, Jennifer Cano, Aitor Lewkowyzc, Michael Mulligan, Charles Rabideau, and Luiz Santos.  OP wishes to acknowledge support from the Simons foundation (\# 385592, Vijay Balasubramanian) through the It from Qubit collaboration. TLH acknowledges the support
of National Science Foundation CAREER Grant No. DMR-1351895 and the ICMT at UIUC.  RGL is supported by the US Department of Energy under contract DE-FG02-13ER42001.

\appendix\numberwithin{equation}{section}

\section{Appendix: Geometric Interpretation of Primitivity}\label{sectGeomPrim}

In this Appendix, we explore some of the implications of the primitivity condition of Section \ref{ClassSympBCs}.  We want to look at this condition \emph{geometrically} and show that it is a sufficient condition for uniquely embedding the torus $U(1)^N\simeq \mathbb T_{\Lambda_0}\hookrightarrow U(1)^{2N}\simeq \mathbb T_\Lambda$.  Recall that primitivity requires that the ${\tiny\left(\begin{array}{c}2N\\N\end{array}\right)}$ $N\times N$ minors of the matrix 
\beq
\mathbb P=\left(\begin{array}{c}v^{(L)}\\-v^{(R)}\end{array}\right)
\eeq
have gcd 1.  The key to this is to note that $\mathbb P$ encodes how many times $\mathbb T_{\Lambda_0}$ wraps $\mathbb T_\Lambda$ as one traverses through one of its cycles.  This is easiest to see in a linear embedding of $\mathbb T_{\Lambda_0}$ into $\mathbb T_\Lambda$.  That is, given global coordinates $\vec\theta=\{\theta_1,\ldots, \theta_N\}$ on $\mathbb T_{\Lambda_0}$ all ranging from $[0,1]$, and coordinates $\{\vec\theta_L;\vec\theta_R\}=\{\theta_{L,1},\ldots, \theta_{L,N};\theta_{R,1},\ldots,\theta_{R,N}\}$ on $\mathbb T_\Lambda$, also ranging from $[0,1]$, then the map $\mathbb P\cdot \vec\theta$ is a topological embedding when $v^{(L)}$ and $v^{(R)}$ are non-degenerate integral matrices.  The submanifold is then the graph of $\{\vec\theta_L,-v^{(R)}{v^{(L)}}^{-1}\cdot \vec\theta_L\}$, or alternatively, $\{-v^{(L)}{v^{(R)}}^{-1}\cdot \vec\theta_R,\vec\theta_R\}$, inside of $\mathbb T_\Lambda$.  However, several choices of $v^{(L)}$ and $v^{(R)}$ yield the same graph.  Let us illustrate this with a simple example.

Let $K^{(L)}=9$ and $K^{(R)}=1$ be 1$\times$1 $K$-matrices.  Ignoring primitivity, solutions to the gluing conditions are $v^{(L)}=m$, $v^{(R)}=\pm 3m$ for some integer $m$.  This embedding of $U(1)\hookrightarrow U(1)^2$ is given in the figure below.  The embedded submanifold in fact depends only on the ratio $-v^{(L)}/v^{(R)}$; different choices of $m$ give different coverings of the same manifold.  Requiring $v^{(L)}$ and $v^{(R)}$ to be relatively prime then fixes an equivalence class of this ratio which corresponds to the minimal covering.\begin{figure}[h!]
  \centering
  \begin{tabular}{ c c c }
    \includegraphics[width=.3\textwidth]{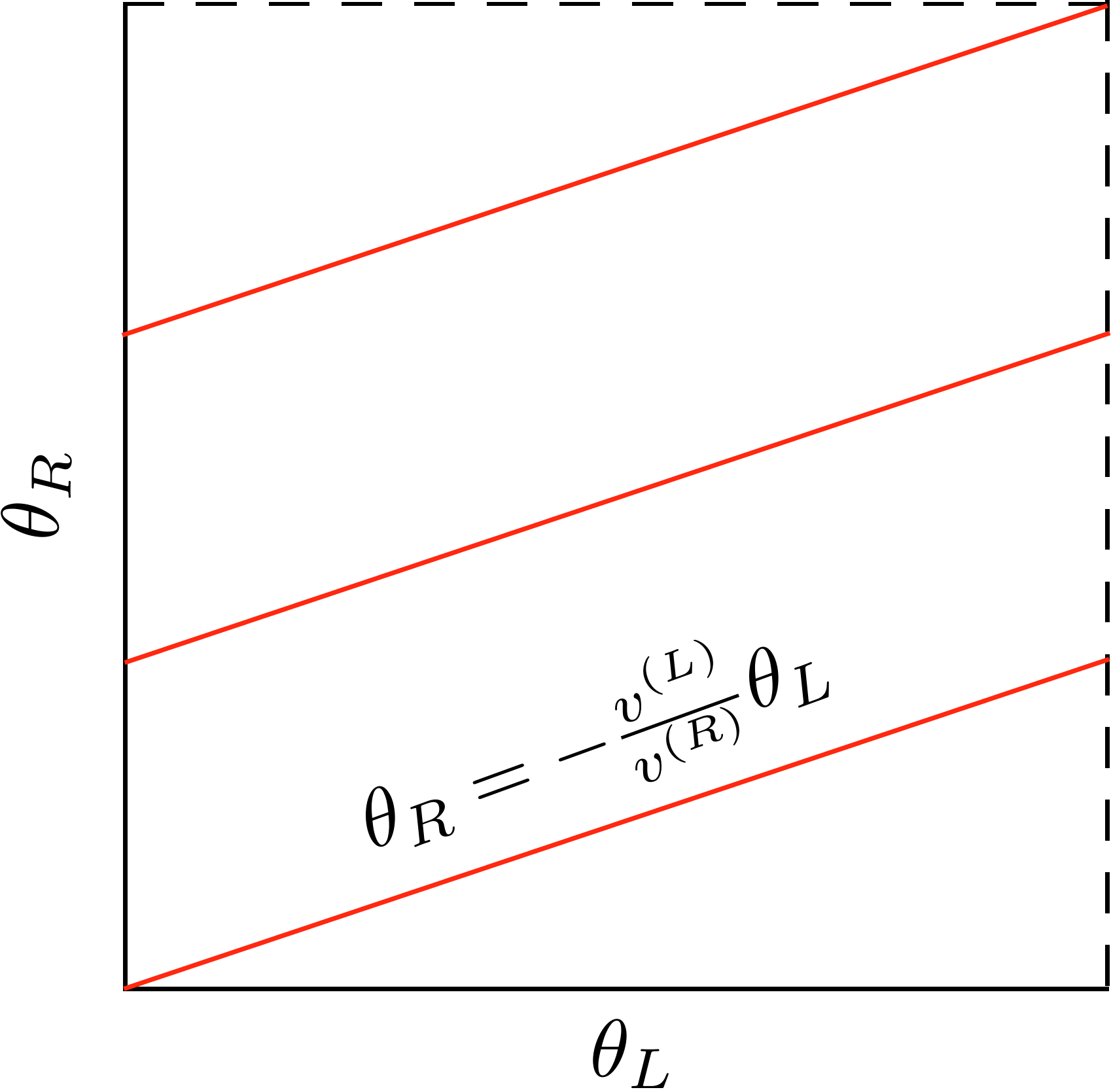} & \hspace{.8 cm} & \includegraphics[width=.5\textwidth]{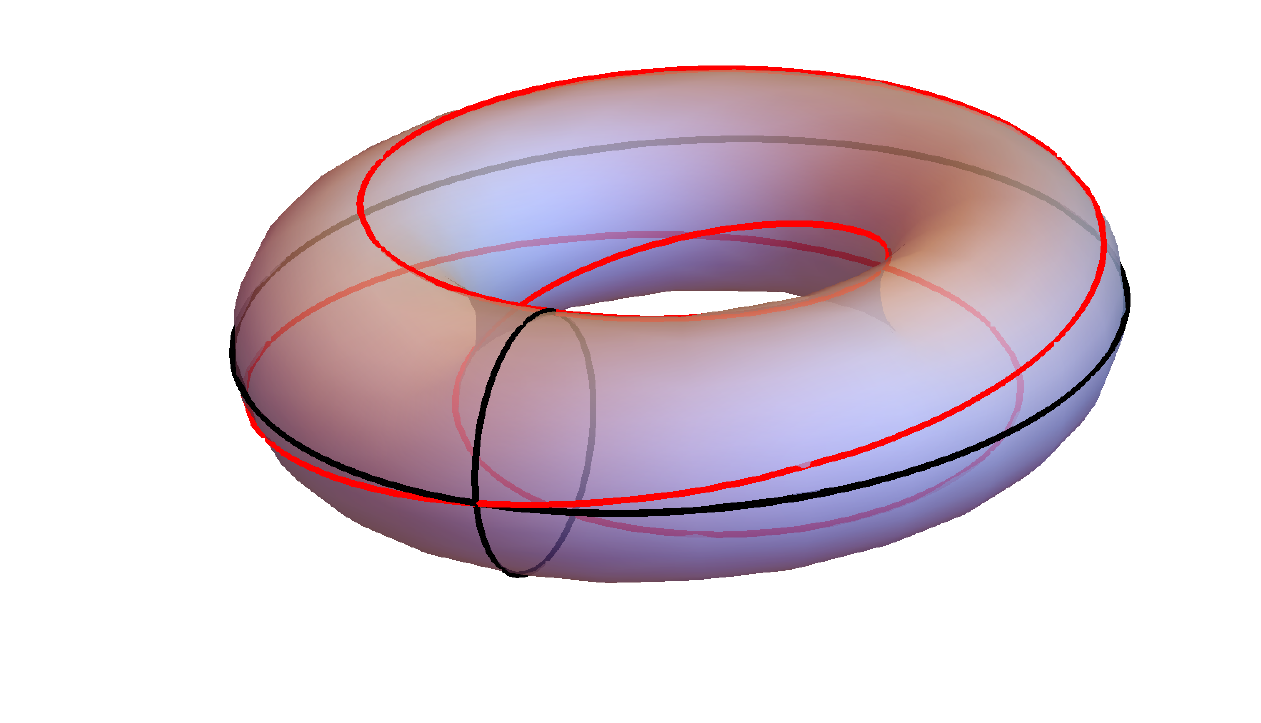}
    \end{tabular}
      \caption{\small{\textsf{(Left) The graph of $v^{(L)}=m$, $v^{(R)}=-3m$.  Note that the value of $m$ is irrelevant for the graph.  (Right) The same graph as wrapped on the torus.}}}
\end{figure}
\\\\
This same principle can be extended to the $U(1)^N$ case.  Let us focus on the graph of $\{\vec\theta_L;\; -v^{(R)} {v^{(L)}}^{-1}\cdot \vec\theta_L\}$.  The key point is that \emph{every element of the matrix $v^{(R)}{v^{(L)}}^{-1}$ is a ratio of two $N\times N$ minors of $\mathbb P$.}  To see this note that for any $N\times N$ invertible matrix, the inverse can be written as
\begin{equation}
{v^{(L)}}^{-1}=\frac{1}{\det v^{(L)}}\bs C_L^T,
\end{equation}
where $\bs C_L^T$ is the matrix of cofactors of $v^{(L)}$ defined by its $(N-1)\times (N-1)$ minors via
\begin{equation}
{\bs C_L}_{ij}=(-1)^{i+j}m^{(L)}_{ij}
\end{equation}
where $m^{(L)}_{ij}$ is the minor formed from removed the $i^{th}$ row and $j^{th}$ column from $v^{(L)}$.  A typical element of $v^{(R)}{v^{(L)}}^{-1}$ is then
\begin{equation}
\left(v^{(R)}{v^{(L)}}^{-1}\right)_{ij}=\frac{1}{\det v^{(L)}}\sum_{k=1}^Nv^{(R)}_{ik}(-1)^{j+k}m_{jk}^{(L)}.
\end{equation}
However, this is the ratio of $\det v^{(L)}$ and the determinant of the matrix formed from replacing the $j^{th}$ row of $v^{(L)}$ with the $i^{th}$ row of $v^{(R)}$, which are both $N\times N$ minors of $\mathbb P$.  The same logic can be run for the equivalent parameterization of the graph as $\{-v^{(L)}\cdot {v^{(R)}}^{-1}\cdot \vec\theta_R;\;\vec\theta_R\}$.  Thus an embedding in which all of the minors share a common factor yields precisely the same graph as one in where they are relatively prime.  Hence, primitivity can be seen as a geometric condition for eliminating equivalent embeddings.

\section{Appendix: Direct Calculation of Homogeneous Chern-Simons Topological Entanglement}\label{appCShomoTEE}
In section \ref{entcutalongint}, we showed through a change of path integral variables that the topological R\'enyi entropy across the heterointerface in $S^3$ can be regarded as the R\'enyi entropy of an effective homogenized theory.  One can imagine then evaluating this homogenized path integral from the standard surgery arguments (e.g. \cite{Witten:1988hf}).  However, without a well defined notion of cutting and gluing along a heterointerface, we supplement this with a more direct evaluation of the path integral on $S^3$.  To be more specific, after performing the Fadeev-Popov procedure, the Chern-Simons path integral on a three manifold $M_3$, can be evaluated as \cite{Blau:1989bq}
\begin{equation}
Z_{M_3}[K]={{\det}'}^{-1/4}_{\Omega_1}\left(\frac{K^2}{16\pi^2}\Delta_1\right){{\det}'}^{-1/4}_{\Omega_3}\left(\frac{K^2}{16\pi^2}\Delta_3\right){{\det}'}_{\Omega_0}\left(\frac{K}{4\pi}\Delta_0\right),
\end{equation}
where $\det'_{\Omega_p}$ is the determinant (excluding zero modes) over the vector space of $p$-forms on $M_3,$ and $\Delta_p=d^\dagger d+d\,d^\dagger$ is the corresponding Laplace operator\footnote{To define $\Delta_p$, we need to introduce a Riemannian metric on $M_3$.  Indeed this metric is introduced in the gauge-fixing stage.  Remarkably, however, the following results are independent of the metric chosen \cite{Blau:1989bq}.} on $\Omega_p$.  We are interested in factoring out the determinant of the $K$-matrix from this product:
\begin{equation}
Z_{M_3}[K]\simeq\left(\det \frac{K}{4\pi}\right)^{\mathscr P_0-\mathscr P_1/2-\mathscr P_3/2}\Big({{\det}'}_{\Omega_1}^{-1/4}\left(\Delta_1\right){{\det}'}_{\Omega_3}^{-1/4}\left(\Delta_3\right){{\det}'}_{\Omega_0}\left(\Delta_0\right)\Big)^N
\end{equation}
where $N$ is the rank of $K$ and the combination of powers $\mathscr P_0-\mathscr P_1/2-\mathscr P_3/2$ will be discussed below.  It is well known that the second factor, i.e., the one raised to the power $N,$ is a topological invariant of $M_3$ related to the Ray-Singer torsion\cite{Ray:1973sb, Schwarz:1978cn}: $T_{R.S.}^{-N/2}$.  Because $T_{R.S.}$ is a probe of the underlying manifold, and not of the anyonic ground state degeneracy (that is to say it is independent of $K$), we disregard this contribution to the path integral (that is, we normalize $Z_{M_3}[K]$ by $Z_{M_3}[\mathbb I_{N\times N}]$).  We regulate the powers $\mathscr P_p$ through zeta function methods and so 
\begin{equation}
\mathscr P_p=\zeta_{\Delta_p}(0)
\end{equation}
for the spectral zeta function of $\Delta_p$ (excluding the zero modes).  We simplify this by noting that $\Omega_p$ admits the Hodge decomposition $\Omega_p=\Omega_p^T\oplus\Omega_p^L\oplus\Omega_p^h$ where $\Omega_p^T$, $\Omega_p^L$, and $\Omega_p^h$ are the transverse, longitudinal, and harmonic eigenspaces of the Laplacian, respectively\footnote{That is any $p$-form can be written as $\omega_p=d^\dagger \sigma_{p+1}+d\eta_{p-1}+\chi^h_p$ for some $p+1$ form $\sigma$, some $p-1$ form $\eta$, and $\chi$ annihilated by the Laplace operator.}.  It can be shown that $\Omega_{p}^L$ is isomorphic to $\Omega_{p-1}^T$ and furthermore, by Hodge duality, that $\Omega_p^{T,L}\simeq\Omega_{d-p}^{L,T}$ \cite{Elizalde:1996nb}.  Excluding the zero modes, then the regulated power of $k$ can be further decomposed into
\begin{equation}
-\frac{1}{2}\left(\mathscr P_1^T+\mathscr P_1^L\right)-\frac{1}{2}\mathscr P_3^L+\mathscr P_0^T=-\frac{1}{2}\left(\mathscr P_1^T+\mathscr P_0^T\right)-\frac{1}{2}\mathscr P_0^T+\mathscr P_0^T=-\frac{1}{2}\mathscr P_1^T
\end{equation}
so we see that only the transverse subspace of the original one-forms contribute to this power, as we should have expected from a gauge-invariant theory.\\
\\
Now let us specialize to $M_3=S^3$.  Using the standard metric on $S^3$, the degeneracies and eigenvalues for the transverse eigenfunctions of $1$-form Laplacian 
are $D_\ell=2\ell(\ell+2)$ and $\lambda_\ell=-(\ell+1)^2$, respectively \cite{Elizalde:1996nb}. The zeta function is then
\begin{align}
\zeta_{-\Delta_1^T}(s)=&\sum_{\ell=1}^\infty2\ell(\ell+2)(\ell+1)^{-2s}=
2\zeta_{H}(2s-2;2)-2\zeta_H(2s;2)
\end{align}
where $\zeta_H(z;q)$ is the Hurwitz zeta function, defined for $\text{Re}(z)>1$ by $\displaystyle\sum_{n=0}^\infty\left(n+q\right)^{-z},$ and then analytically continued for complex $z$.  The special values of $\zeta_H(z;q)$ are well known and now we may take the $s\rightarrow 0$ limit.  In particular, we note $\zeta_H(0;q)=\frac{1}{2}-q$ and $\zeta_H(-n; q)=-\frac{B_{n+1}(q)}{n+1}$ where $B_{n+1}(q)$ is the Bernoulli polynomial when $n$ is a natural number.  Doing so we have
\begin{equation}
\mathscr P_1^T=2\zeta_H(-2;2)-2\zeta_H(0;2)=1.
\end{equation}
The end result is that
\begin{equation}\label{HomoAns}
Z_{S^3}[K]\simeq{\det}^{-1/2} K
\end{equation}
consistent with the known topological entanglement entropy.\\\\
As a brief aside, we can convince ourselves that these path integrals are properly normalized by repeating this procedure on $M_3=S^1\times T^2$.  We then expect that the answer should be the dimension of the Hilbert space on the torus: $Z_{S^1\times T^2}=\text{dim}\mathcal H_{T^2}=\det K$.  Indeed by explicit calculation using the flat product metric on $S^1\times T^2\simeq S^1\times S^1\times S^1$, the eigenvalues for $\Delta_1^T$ are
\begin{equation}
\lambda^2=\frac{1}{4}\left(m^2+n^2+p^2\right)
\end{equation}
where $m,n,$ and $p$ are integers, not all simultaneously zero.  Each non-zero eigenvalue has degeneracy 2 and so 
the spectral zeta function for this Laplacian is then
\begin{align}
\zeta_{\Delta_1^T}(s)=\sum_{m,n,p\in\mathbb Z^3\setminus\{0,0,0\}}2\left(\frac{m^2+n^2+p^2}{4}\right)^{-s}=2^{2s+1}\zeta_{E,3}(s;0)
\end{align}
where $\zeta_{E,D}(s;q)=\sum_{\vec n\in \mathbb Z^D\setminus\vec 0}\left(|\vec n|^2+q^2\right)^{-s}$ converges for $s>D/2$ to the Epstein zeta function.  
Analytically continuing this to $\zeta_{E,3}(0;0)=-1$ see then that the standard result is exactly reproduced:
\begin{equation}
Z_{S^1\times T^2}\simeq \det K =\text{dim}\mathcal H_{T^2}.
\end{equation}

\providecommand{\href}[2]{#2}\begingroup\raggedright\endgroup


\end{document}